\documentclass[twoside,a4paper]{JHEP3}
\usepackage[dvips]{graphicx}
\usepackage{amsmath}
\usepackage{cite}
\usepackage{epsfig}
\usepackage[english]{babel}
\usepackage[latin1]{inputenc}
\usepackage[T1]{fontenc}
\usepackage{amsfonts}
\usepackage{float}
\usepackage{caption}
\newcommand{\be}{\begin{equation}}
\newcommand{\ee}{\end{equation}}
\newcommand{\bea}{\begin{eqnarray}}
\newcommand{\eea}{\end{eqnarray}}

\title{Metastable SUSY Breaking, de Sitter Moduli Stabilisation and
  K\"ahler Moduli  Inflation}
\author{Sven Krippendorf$^1$, Fernando Quevedo$^{1,2}$
\\ $^1$ DAMTP, Centre for Mathematical Sciences,\\ $\,$ Wilberforce Road, Cambridge, CB3 0WA, United Kingdom
\\ $^2$ CERN PH-TH, CH 1211, Geneva 23, Switzerland.}
\preprint{DAMTP-2009-12\\
CERN-PH-TH/2009-014}
\abstract{
We study the influence of  anomalous $U(1)$ symmetries and their
associated D-terms on the vacuum structure of global field theories once
they are coupled to ${\cal N}=1$  supergravity and in the context of
string compactifications with  moduli stabilisation.
In particular, we focus on a IIB string motivated construction of the ISS
scenario and examine the influence of one additional $U(1)$ symmetry on
the vacuum structure. We point out that in the simplest one-K\"ahler
modulus compactification, the original ISS vacuum gets generically
destabilised by a runaway behaviour of the potential in the modulus
direction. In more general compactifications with several K\"ahler moduli,
we find a novel realisation of the LARGE volume scenario with D-term
uplifting to de Sitter space and both D-term and F-term supersymmetry
breaking. The structure of soft supersymmetry breaking terms is determined
in the preferred scenario where the standard model cycle is
not stabilised non-perturbatively and found to be flavour universal. Our scenario  also provides a purely
supersymmetric realisation of K\"ahler moduli (blow-up and fibre)
inflation, with similar observational properties as the original
proposals but without the need to include an extra (non-SUSY) uplifting term.
}
\begin{document}
\section{Introduction}
In this article we address the following issues related to supersymmetry breaking, moduli stabilisation, de Sitter vacua and moduli inflation in string compactifications:
\begin{enumerate}
\item{}
Despite the recent great success on moduli stabilisation in string theory, the value of the vacuum energy after moduli stabilisation naturally corresponds to anti de Sitter space. This is understandable from the effective field theory point of view due to the fact that the scalar potential of ${\cal N}=1$ supergravity is not positive definite and therefore the local minima tend to be at negative values of the vacuum energy. The original uplifting mechanism to de Sitter space proposed in reference \cite{Kachru:2003aw} by introducing anti D3 branes
requires the explicit breaking of supersymmetry in the effective field
theory. Further uplifting mechanisms have been proposed
\cite{Burgess:2003ic,Saltman:2004sn}. Despite partial success, there is not at the
moment a compelling mechanism  for de Sitter uplifting. The consideration of
D-terms as proposed in \cite{Burgess:2003ic} has to be implemented in concrete setups
since, even though
D-terms add a positive definite
contribution to the effective potential, it is known that if the
F-terms vanish then D-terms also vanish
\cite{dterms}. Concrete examples have been provided in field
theoretical \cite{Achucarro:2006zf} and string inspired models \cite{Cremades:2007ig,Haack:2006cy}.

\item{}
In the past two years there has been a large amount of work on the natural appearance of metastable supersymmetry breaking vacua in global supersymmetry and in particular, the breakdown of global supersymmetry can be achieved significantly simpler. This is the ISS scenario \cite{Intriligator:2006dd}. Some realisations of this mechanism in
string models have also been  obtained \cite{braneconstructions}.
Nevertheless, there is an implicit assumption in this mechanism, $i.e.$ that the gauge coupling of the corresponding gauge theory is constant and
therefore the dynamical non-perturbative scale $\Lambda\sim
e^{-a/g^2}$ appears as a constant in the effective action.
In string theory, however, the
gauge coupling is field dependent
$1/g^2 \sim {\rm Re}\, T$ with $T$ the complex scalar component of a
chiral superfield corresponding to a closed string modulus. Therefore the  exponential
dependence of $\Lambda$ would tend to give a runaway behaviour to the scalar potential as a function of ${\rm Re}\, T$.

\item{}
In the past few years several string theory mechanisms for
cosmological inflation have been proposed. The inflaton field
corresponds to either an open string mode as in brane-antibrane
\cite{braneantibrane}, D3/D7 \cite{d3d7}, monodromy \cite{Silverstein:2008sg} or Wilson line \cite{wilsonline} inflation or a closed string modulus as in Racetrack \cite{BlancoPillado:2004ns, BlancoPillado:2006he}, K\"ahler moduli \cite{Conlon:2005jm, Bond:2006nc}, monodromy \cite{McAllister:2008hb} or fibre \cite{Cicoli:2008gp} inflation. In most of
these scenarios, the realisation of inflation depends crucially on
the uplifting mechanism for de Sitter moduli stabilisation.
Therefore it tends to go beyond the ${\cal N}=1$ supersymmetric
effective action if the uplifting mechanism is the presence of anti
D3 branes as in KKLT. It would be desirable to find a string
inflation mechanism derived from a fully ${\cal N}=1$ supersymmetric
action. Recently in \cite{Covi:2008cn}, general constraints on this
possibility have been found, substantially restricting the class of
supersymmetric models that can give rise to inflation. It is a
challenge to find a concrete stringy realisation of inflation that
satisfies those constraints.
\end{enumerate}
We address these three issues by considering a class of models that can be realised in
terms of D-brane orientifold constructions in fluxed Calabi-Yau
manifolds.  In particular we consider a system of magnetised D7 branes with chiral matter fields.
 The existence of anomalous $U(1)$'s induces
Fayet-Iliopoulos D-terms. Fluxes of RR form give rise to a tunable constant
term in the superpotential and also the presence of matter fields provides
a non-perturbative superpotential. Depending on the number of D-branes this can be realised in
the electric or magnetic phase of SQCD.
The electric phase was considered in \cite{Cremades:2007ig}. Here we will concentrate on the magnetic phase.
This makes contact with the ISS models \cite{Intriligator:2006dd}
and then naturally addresses the issue of the runaway behaviour of the scalar potential.
We find that in the simplest case of one single K\"ahler
modulus, instead of a metastable vacuum, the potential runs away in the direction of the modulus
corresponding to the size of the 4-cycle wrapped
by the magnetised D7 branes, ruining the interesting properties of the ISS mechanism.
However, once we have several K\"ahler moduli, the situation changes drastically.
We find minima of the scalar potential at finite values of the fields
with supersymmetry broken as in the ISS mechanism. Furthermore, depending on the values of the free parameters,
the minima correspond to either anti de
Sitter or de Sitter spaces. In the latter case, it provides a natural uplifting mechanism for moduli stabilisation,
addressing also the first issue
mentioned above.

Even though the presence of matter fields and D-terms changes the
structure of the scalar potential very much, we find remarkably
that, once the matter fields are integrated out, the scalar
potential for moduli fields ends up to be very similar, but not
identical,  to the one found in \cite{Balasubramanian:2005zx,
Conlon:2005ki} and provides a new realisation of the LARGE volume
 scenario of moduli stabilisation.

Due to the similarity with the original LARGE volume scenario, we
derive the structure of soft supersymmetry breaking terms and
compare with previous results, we then revisit the K\"ahler moduli
and fibre inflation scenario in our setup and find that they are
also naturally realised, therefore providing a pure ${\cal N}=1$
realisation of these inflationary scenarios.

\section{The ISS model and its embedding into String Theory}
We want to discuss the vacuum structure of the magnetic dual within the Seiberg duality setup of SQCD which arises in the range $N_C<N_F<3/2 N_C.$ As commonly known in this phase of SQCD there exists a dual description of the low-energy field theory in terms of an infrared-free modified version of SQCD, the so-called free magnetic dual description. It is characterised by the meson fields $\Phi,$ quarks $q$ and anti-quarks $p$ which transform under the following symmetry groups:
 \begin{equation}
\begin{matrix}
& SU(N_F-N_C) & SU(N_F)_L & SU(N_F)_R & U(1)_B & U(1)_A & U(1)_R\cr
 q & \Box & \Box& 1 &1&1&0\cr
p&\bar{\Box}&1&\bar{\Box}&-1&1&0\cr
\Phi&1&\bar{\Box}&\Box&0&-2&2\cr
\end{matrix}
\end{equation}
where the only gauge symmetry is the $SU(N_F-N_C)$ symmetry. The K\"ahler potential is taken to be canonical and looks like
\begin{equation}
K=|q|^2+|p|^2+|\Phi|^2\, .
\end{equation}
In order to obtain dynamical SUSY breaking one introduces a mass term. Hence the most general invariant superpotential we can write down is
\begin{equation}
W=\Lambda \frac{q\Phi p}{\mu}+\Lambda m \Phi\, ,
\end{equation}
where $m$ denotes the mass parameter of our theory, $\Lambda$ the dynamical scale of our theory, and $\mu$ is
another parameter that is determined by the duality relations. In the above equations and in the whole article
we omit indices where possible. Please note that the scale $\Lambda$ has to be introduced due to dimensional
requirements.

The metastable vacuum solution is given by $\Phi=0$ and $p=q=i \sqrt{\mu m}.$ A derivation can be found in
\cite{Intriligator:2007py}. The mass term explicitly breaks the $U(1)_A$ global symmetry.\\

There have been several realisations of the ISS scenario within
string theory in terms of different brane configurations
\cite{braneconstructions}. Here we are more interested in capturing the main
ingredients that a string theory realisation will add to this
scenario, namely the fact that the parameters $m$ and $\Lambda$ have
to be dynamical variables. In particular $\Lambda$ in string theory
having a non-perturbative origin is of the form $\Lambda\sim
e^{-aT}$ with $T$ a closed string modulus. One simple explicit
realisation is the magnetic version of the case considered in
\cite{Cremades:2007ig}. The basic setup is an orientifold model
consisting of a stack with $N_C+1$ branes, from which one has a
non-vanishing magnetic flux leading to $U(N_c)\times U(1)$ gauge
theory. The open strings going from the stack of $N_c$ branes to the
magnetised  one are the elementary chiral fields $Q$. The
anti-chiral fields $\tilde{Q}$ have their end-points in the stack of
$N_c$ branes and the  orientifold image of the magnetised brane,
whereas the fields $\rho$ with endpoints in both images of the magnetised
brane are singlets under the non-abelian group but charged
under the $U(1)$. The effective field theory in the electric phase
($N_F<N_c$) was studied in detail in \cite{Cremades:2007ig}.

Here we will consider the magnetic phase appropriate for
$N_c<N_F<3N_c/2$ for which instead of $Q,$ $\tilde{Q}$ the fundamental
fields are the dual fields which we denote $q,$ $p$ and the meson-like
field $\Phi$ . The singlet field $\rho$ will play the role of the
mass parameter. Since generically the $U(1)$ is anomalous, the
modulus $T$ whose real part is the inverse gauge coupling is also
charged under the $U(1)$. The $U(1)$ charges are given in the table.
\begin{center}
\begin{tabular}[h]{c | c | c | c | c | c }
 &  $q$ &$p$ & $\rho$ & $\Phi $& $e^{aT}$ \\ \hline
$U(1)$ & $-1/2$ & $-1/2$ & $-1$ & $2$ & $1$\\
\end{tabular}
\end{center}
The effective field theory will be determined by the superpotential $W$, the gauge kinetic function $f$ and K\"ahler potential $K$ as follows: the gauge kinetic function for the gauge fields on the relevant D7 brane is at leading order $f=T_s$, where $T_s$ is the K\"ahler modulus whose real part is $\tau_s$. The superpotential is of
the form
\be W=W_0 + W_{\rm np} \ee
with $W_0$ a flux induced superpotential which is a constant after fixing the complex structure moduli and the non-perturbative superpotential is taken to be of the 
moduli dependent ISS form \footnote{A full string theory derivation of this non-perturbative superpotential is not yet available 
(see \cite{Nakayama:2007du} for a previous discussion of this system).}:
\be W_{\rm np}=  \alpha e^{-a T_s}\left(\frac{p\Phi q}{\mu}+\Phi \rho\right)\, ,
\ee
 Here $\alpha, a, \mu$ are constants.
The K\"ahler potential is
\be
K= -2 \log\left(\cal{V}+\xi\right) + K_{\rm matter}\, , \ee
where $\cal{V}$ is the volume written in terms of the K\"ahler
moduli $\tau_a$ and $\xi$ corresponds to the leading order $\alpha'$-corrections. The chiral
matter K\"ahler potential is only known in a small field expansion as a function of the K\"ahler moduli, with the complex structure dependence unknown, however this is enough for our purposes here. From the analysis in \cite{Conlon:2006tj} we can write:
\be
K_{\rm matter}=
\frac{\tau_s^n}{{\cal{V}}^{2/3}}\left(|\Phi|^2+|\rho|^2+|q|^2+|p|^2\right)\,
, \ee
where $n$ is the modular weight.
 We can then ask how the structure of the vacuum for the ISS scenario is modified once the moduli dependence of the superpotential and K\"ahler potential are included.

\section{The One-modulus Case}
We first consider the simplest case of a one K\"ahler modulus Calabi-Yau with  $T=\tau+i b,$ where $\tau$ denotes the Einstein frame volume of the 4-cycle $X$ and $b=\int_X C_4.$ This is similar to the setup presented by Nakayama et al. \cite{Nakayama:2007du} and is considered as a stabilisation in the spirit of KKLT \cite{Kachru:2003aw}. After discussing the setup, we will show why it is not possible to stabilise the K\"ahler moduli and brane moduli in a viable regime.

The K\"ahler potential takes the simple form (here the explicit dependence on $g_s$ and $M_P$ is included since they are important for the numerical estimates):
\begin{equation}
K=-2 M_P^2 \log \left({\cal V}+\frac{\xi}{g_s^{3/2}}\right)+\frac{1}{(g_s \tau)^{n}}\left(|\Phi|^2+|\rho|^2+|q|^2+|p|^2\right)\, ,
\end{equation}
where we kept the modulus weight general and included the leading order $\alpha'-$corrections. The volume simply is assumed to be ${\cal V}=\tau_1^{3/2}$ and is in the Einstein frame. The matter fields have mass dimension 1. The superpotential becomes
\begin{equation}
W=g_s^{3/2} M_P^3 W_0+M_P g_s^{4/3} e^{-aT}\alpha \left(\frac{p\Phi q}{\mu}+\rho \Phi \right)\, .
\end{equation}
We can now calculate the D-term potential for the anomalous $U(1)$  with the previously discussed charge assignments. Setting $g_s=M_P=1$, for simplicity of notation,
 the D-term potential becomes:
\begin{multline}
V_D=\frac{1}{2\tau} \left(\frac{- |\rho|^2 - \frac{1}{2}(|q|^2+|p|^2)+ 2 |\Phi|^2}{\tau^{n}}+\frac{1}{a}\left(\frac{-2\sqrt{\tau}}{3(\xi+\tau^{3/2})}\right. \right.
 \\ \left. \left.-n \frac{|\rho|^2+|p|^2+|q|^2+|\Phi|^2}{2  \tau^{n+1}}\right)\right)^2,
\end{multline}
The F-term potential is calculated from the standard supergravity formula
\begin{equation}
V_F=e^{\frac{K}{M_P^2}}\left(D_i W D_{\bar{j}} \bar{W} K^{-1}_{i\bar{j}}-3 \frac{|W|^2}{M_P^2}\right).
\end{equation}
with $D_iW=\partial_iW+ W \partial_i K/M_P^2$.
If we assume no implicit dependence of the matter fields on the volume, the leading order contributions to the F-term potential are given by:
\begin{eqnarray}
e^{K}&\sim&\frac{1}{ \left(\tau^{3/2}+\xi\right)^2}\, ,\\
\partial W \partial \bar{W} K^{-1}&\sim&\left(\frac{4}{3}\tau^2 - \frac{4 n\tau^{2-n}}{9}\right)a^2 e^{-2 a \tau}\alpha^2 \left|\frac{p\Phi q}{\mu}+\rho \Phi\right|^2\,
 \text{ for }n>0\, ,\\
W \partial W \partial \bar{K} K^{-1}+ \text{ c.c.}&\sim & 4 W_0  a \alpha\, \tau \text{Re}\left(e^{-a T}\left(\frac{p\Phi q}{\mu}+\rho \Phi\right)\right), \\
\label{dkdk1} |W|^2\partial K \partial \bar{K} K^{-1}&\sim&|W_0|^2 \left(3+\frac{3 \xi}{2 \tau^{3/2}}+(1-n)\frac{|p|^2+|q|^2+|\rho|^2+|\Phi|^2}{\tau^{n}}\right).
\end{eqnarray}
 Bearing in mind that the flux parameter  $W_0$ has to be exponentially small in KKLT-like setups,
 the key observation for the following discussion is the fact that we have got a hierarchy between the D-term potential and the F-term potential, simply since the
F-term potential is exponentially suppressed and the D-term potential is not. Therefore the D-term potential has to vanish almost exactly, that is it should
vanish up to the exponential suppression of the F-term potential.

Because of the D-term dominance, we are required to address a completely different stabilisation procedure compared to the global field theory setup of ISS.
It is actually not
possible to integrate out the fields at the minimum of the global SUSY model.

The multi-field minimisation process can be simplified (as in the global case) by noting that
the D-term potential arising
from the $SU(N_F-N_C)$ gauge theory can be minimised by simply requiring the VEV of both type of fields $q$ and $p$ to be the same, since:
\begin{equation}
V_D^{SU(N_F-N_C)}=\frac{g^2}{2}\sum_A \left(\text{Tr } q^\dagger T_A q-\text{Tr } pT_A p^\dagger\right)^2\; ,
\end{equation}
where $T_A$ denotes the generator of the fundamental representation of $SU(N_F-N_C).$
It can further be seen that the minimum corresponds when both vanish $q=p=0$.\footnote{See Appendix C for a detailed discussion.} Since the D-term is dominant the remaining matter fields
are related by a condition of the form  $|\rho|^2\sim |\Phi|^2-b,$ where $b$ is determined by the leading order FI term.
This reduces the potential as a function of, say, $|\Phi|$ and $\tau$. As expected we do not find a minimum for large values
of $\tau$ where the potential shows the standard runaway behaviour as illustrated in the figure
 \ref{runaway} below.\\
 \begin{center}
\begin{minipage}{0.7\textwidth}
\centering
\includegraphics[width=0.95\textwidth]{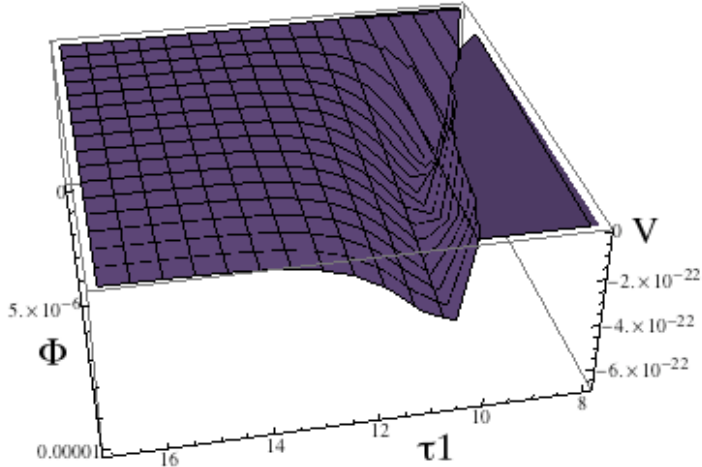}
\end{minipage}\hfill
\begin{minipage}{0.3\textwidth}
\captionof{figure}{\footnotesize{This plot shows the F-term potential in dependence of the K\"ahler modulus $\tau$ and the remaining matter degree of freedom $\Phi.$ We
display small values of $\Phi$ in order to show the reader the runaway behaviour, which is clearly shown in the picture. Keeping $\Phi$ constant would give the typical KKLT
 situation.}\label{runaway}}
\end{minipage}
\end{center}
For small values of $\tau$ local minima can be found, but in a regime that does not justify keeping only the leading order corrections to $K$ \cite{Nakayama:2007du}.
\section{The Two-moduli Case}

Since the one-modulus case leads naturally to the runaway behaviour, it is tempting to conclude that string theory
realisations of the ISS mechanism do not have a metastable state but a runaway potential. We will now explore the Swiss cheese,
two-moduli case which has proven to lead to qualitatively different vacuum structure in the
LARGE volume scenario of moduli stabilisation and reconsider moduli stabilisation in our set-up.

We will then consider the geometry of the Calabi-Yau defined as the surface in projective space $\mathbb{P}_{[1,1,1,6,9]}$ where the volume is given by
\begin{equation}
{\cal V}=\tau_1^{3/2}-\tau_2^{3/2}\, ,
\end{equation}
where $\tau_1$ denotes the large 4-cycle and $\tau_2$ the small one. Then our stringy setup is given by:
\begin{eqnarray}
K&=&-2M_P^2\log{\left(\tau_1^{3/2}-\tau_2^{3/2}+\frac{\xi}{g_s^{3/2}}\right)}+\frac{g_s^n\tau_2^{n}}{g_s\tau_1}(|p|^2+|q|^2+|\rho|^2+|\Phi|^2)\, ,\\
W&=&M_P^3 g_s^{3/2}W_0+\alpha M_P g_s^{1/2+n}e^{-a T_2} \left(\frac{p\Phi q}{\mu}+\Phi \rho\right)\, .
\end{eqnarray}
In the following we will use the charge assignment
\begin{center}
\begin{tabular}[h]{c | c | c | c | c | c }
 &  $q$ &$p$ & $\rho$ & $\Phi $& $e^{aT_2}$ \\ \hline
$U(1)$ & $1$ & $1$ & $2$ & $-1$ & $1$\\
\end{tabular}
\end{center}
That is the D7 branes are wrapping the small four-cycle and then it is $T_2$ that gets charged under $U(1)$ and not $T_1$.
Then the D-term potential becomes
\begin{multline}
\nonumber V_D=\frac{g_s}{2 \tau_2}\left[\frac{2|\rho|^2 \tau_2^{n}}{g_s^{1-n}\tau_1}+\frac{(|q|^2+|p|^2)\tau_2^{n}}{g_s^{1-n}\tau_1}-\frac{ |\Phi|^2 \tau_2^{n}}{g_s^{1-n}\tau_1}+\right.\\
\left.+\frac{1}{a}\left(\frac{3 M_P^2\sqrt{\tau_2}}{2\left({\cal V}+\frac{\xi}{g_s^{3/2}}\right)}+\frac{(|\rho|^2+|p|^2+|q|^2+|\Phi|^2) n\tau_2^{n-1}}{2g_s^{1-n}\tau_1}\right)\right]^2 ,
\end{multline}
where the overall $g_s$ arises due to the transformation from string to Einstein frame. In general there are two possible charge assignments; one is displayed above, and the other interchanges the charges $\rho$ and $\Phi$, meaning that both fields are changing their role in the D-term stabilisation.

The F-term potential is given by the usual supergravity formula (divided by $g_s^{2}$ which arises by integrating out the dilaton dependence)
\begin{eqnarray}
V_F&=&\frac{1}{g_s^2}e^{\frac{K}{M_P^2}}\left(D_i W D_{\bar{j}} \bar{W} K^{i \bar{j}}-3 \frac{|W|^2}{M_P^2}\right)\\
\nonumber&=&\frac{e^{\frac{K}{M_P^2}}}{g_s^2}\left(\partial_i W \partial_{\bar{j}} \bar{W} K^{i \bar{j}}+\partial_i W \partial_{\bar{j}} \bar{K} K^{i \bar{j}}\frac{\bar{W}}{M_P^2}+\frac{W}{M_P^2}  \partial_i K \partial_{\bar{j}} \bar{W} K^{i \bar{j}}+\frac{|W|^2}{M_P^4}\partial_i K \partial_{\bar{j}} \bar{K} K^{i \bar{j}}-3 \frac{|W|^2}{M_P^2}\right),
\end{eqnarray}
where the indices run over the two K\"ahler moduli $T_1,T_2$ and matter fields $p,q,\Phi$.
\subsection{Strategy}
Our aim is now to find a minimum of the whole potential at large volume to guarantee stability towards unknown higher order corrections. Therefore we should always keep track of the
power dependence in $\tau_1$ which corresponds directly to the power suppression by the volume. The results presented are only reliable up to some order in the volume
suppression.

Let us start by emphasising the following points:
We can see that the D-term potential is suppressed by $1/\tau_1^2$ whereas the F-term potential is at least suppressed by $1/\tau_1^3.$
This directly implies that the D-term contribution is a priori leading compared to the F-term potential and hence the global SUSY analysis of ISS will be modified.
Also, minimisation with respect to the matter fields will lead to a moduli dependence of their VEVs.

Minimising the whole potential analytically is not possible due to the  complexity of the system corresponding to a potential as a function of six complex fields.
 Instead, we will use the suppression with respect to the volume of every single term as a natural order criterion, since this allows us to discard various higher order
contributions. Our approach can be summarised as follows:
\begin{enumerate}
\item The first step of our minimisation procedure will be to find the
 dependence of the matter fields (e.g. $\chi$) on the large K\"ahler modulus, at leading order, writing $\chi = \tilde{\chi}/{\cal{V}}^m$ and determine $m$
for each matter field $\chi$,
 still not fixing the matter fields completely (leaving $\tilde{\chi}$ still unfixed).
\item After that we can minimise the matter fields completely by fixing $\tilde{\chi}$.
We then end up with a scenario which looks roughly like the original LARGE volume scenario with an additional uplifting D-term.
\item We finally stabilise the potential with respect to the K\"ahler moduli numerically.
\end{enumerate}
\subsection{Matter field stabilisation}
Similar to the one-field case, the fields $p$ and $q$ can be stabilised at $p=q=0$.\footnote{See Appendix C for a detailed discussion.} Then we need to determine $\Phi$ and $\rho$ as functions of the volume.
 We should note at this point that it is not possible to set any of these two fields to zero, in particular $\rho\neq 0,$ since then the contribution from
 $\partial W \partial K K^{-1}$ vanishes as well, and we remain with only positive definite terms in the F-term potential.

We would now like to fix the implicit dependence of the matter fields on the K\"ahler moduli. In order to ensure that there will not be any further leading
order correction to this assumption it is necessary to satisfy one of the following conditions:\\
After fixing the quark and antiquark fields the D-term potential looks like:
\be
V_D=\frac{g_s}{2\tau_2 {\cal V}^{4/3}}\left(\frac{|\rho|^2 \left(36a \tau_2+n\right)}{18a g_s^{1-n}\tau_2^{1-n}}-\frac{|\Phi|^2(18a\tau_2-n)}{18a g_s^{1-n}\tau_2^{1-n}}+\frac{3 M_P^2\sqrt{\tau_2}}{2a {\cal V}^{1/3}}\frac{1}{1+\frac{\xi}{ g_s^{3/2}{\cal V}}}\right)^2\, .
\ee
 Starting from the observation that the only negative contribution in the D-term potential comes from the term including $\Phi$ fields, it is natural to think that this term will cancel the leading order FI contribution.\footnote{The next to leading order FI contribution is of sub-leading order even compared to the leading order F-term potential.} Cancelling the leading order FI-term with the $\Phi$ field implies the following implicit volume dependence:
\begin{equation}
|\Phi|^2=\frac{|\tilde{\Phi}|^2}{ {\cal V}^{1/3}}\, .
\end{equation}
With this assumption we can rewrite the D-term as follows:
\begin{eqnarray}
\nonumber V_D&=&\frac{g_s}{2\tau_2 {\cal V}^{4/3}}\left(\left(\frac{|\rho|^2 \left(36a \tau_2+n\right)}{18a g_s^{1-n}\tau_2^{1-n}}\right)^2+\frac{1}{{\cal V}^{2/3}}\left(\frac{3 \sqrt{\tau_2}}{2a }\frac{M_P^2}{1+\frac{\xi}{ g_s^{3/2}{\cal V}}}-\frac{|\tilde{\Phi}|^2(18a \tau_2-n)}{18a g_s^{1-n}\tau_2^{1-n}}\right)^2\right.\\
&+&\left.  \frac{2}{{\cal V}^{1/3}}\left(\frac{3 \sqrt{\tau_2}}{2a }\frac{M_P^2}{1+\frac{\xi}{ g_s^{3/2}{\cal V}}}-\frac{|\tilde{\Phi}|^2(18a\tau_2-n)}{18a g_s^{1-n}\tau_2^{1-n}}\right) \left(\frac{|\rho|^2 \left(36a \tau_2+n\right)}{18a g_s^{1-n}\tau_2^{1-n}}\right)\right)\, .
\end{eqnarray}
To cancel the FI-term we find:
\begin{equation}
|\Phi|^2=\frac{|\tilde{\Phi}|^2}{ {\cal V}^{1/3}}=\frac{54 M_P^2 g_s^{1-n}\tau_2^{3/2-n}}{2  (18 a\tau_2-n){\cal V}^{1/3}}\overset{a \tau_2\gg 1}{=}\frac{3M_P^2 g_s^{1-n}\tau_2^{1/2-n}}{2 a {\cal V}^{1/3}}\, .
\end{equation}
This assignment minimises the potential with respect to $\Phi$ at leading order. Higher order corrections are in principle of importance if we stabilise $\rho$ at a higher order compared to $\Phi.$ However, we show in the appendix that we can neglect these higher order corrections at our minimum.

Focusing on the mass field $\rho,$ the first guess might be that there should be no difference in the implicit dependence between $\Phi$ and $\rho$ since they occur completely symmetrically after setting the quark fields to zero apart from their appearance in the D-term. In this case, we end up with the following two possibilities:
\begin{enumerate}
\item $\Phi$ and $\rho$ both get VEVs that are of the same order as the FI contribution but we have to cancel the D-term to leading order, so we do have to cancel the FI-term with $\Phi$. This fixes one of the matter fields up to a phase. We then have to minimise the F-term potential on its own but with the following constraint $|\rho|^2\sim|\Phi|^2-b,$ where $b$ is determined by the leading order FI part. The structure of the F-term looks promising since we find the exact same volume suppression as in the LARGE volume scenario and we can minimise with respect to the volume. But when we then try to stabilise the remaining matter field, it turns out that one cannot achieve it at large values for the volume. A detailed proof of this statement can be found in appendix A.
\item $\Phi$ and $\rho$ get VEVs that are larger then the FI-term and they then differ at that "sub-leading" order. The problem now is that the matter fields become so massive that we cannot trust our expansion anymore since we have to consider trans-Planckian dynamics. In addition, one runs into similar problems as above.
\end{enumerate}
Since the D-term potential should be suppressed higher than the F-term potential we would like $\rho$ to be higher suppressed than $\Phi.$ Hence we have to give $\rho$ a higher implicit suppression with respect to the K\"ahler moduli than to $\Phi.$ To find the implicit suppression and have a maningful approximation,
 we demand that we have a very large volume and then, after the full stabilisation process is finished, check if the assumption is consistent. 
This gives us which term should be leading order in $\tau_1,$ namely:
\begin{equation}
\partial W\partial \bar{W} K^{-1} e^K.
\end{equation}
Since the next to leading order contribution from the FI-term is clearly subleading, we have to suppress the term proportional to $\rho$ at least more than the leading order F-term potential term. This constraint implies that the leading order F-term contribution from the term above is independent of $\rho.$ Clearly $\rho$ will occur in the leading order term coming from $\partial W\partial \bar{K} K^{-1} e^K.$ With this implication we can determine the leading order F-term potential which looks as follows:
\begin{multline}
V_F=\frac{M_P^4}{g_s^2\tau_1^3}\left(\frac{g_s^{2+n}|\tilde{\Phi}|^2 \alpha^2 \sqrt{\tau_1} e^{-2a\tau_2}}{M_P^2\tau_2^{n}}+\frac{4a g_s^{2+n}W_0 \alpha \tau_2 e^{-a\tau_2}\text{Re }(\tilde{\rho}\tilde{\Phi})}{M_P^2 \tau_1^{1/4+m}}\right. \\ \left.+\frac{3 W_0^2 \left(\frac{\xi}{g_s^{3/2}}+\frac{2n\tau_2^n|\tilde{\Phi}|^2}{g_s^{1-n}M_P^2} \left(1-\frac{\xi}{g_s^{3/2}\tau_2^{3/2}}\right)\right)}{2 \tau_1^{3/2}}\right),
\end{multline}
where we kept the explicit dependence on $\tilde{\Phi}$ and $m$ reflects the uncertainty in the implicit suppression of $\tilde{\rho}$ with respect to the large K\"ahler modulus and worked in the limit $a\tau_2\gg 1$ for clarity. Terms including higher order corrections in $\Phi$  have not been written down but would not enter at leading order. From the structure above, we can see that we cannot stabilise the potential with respect to $\tilde{\rho}$ with a vanishing D-term since the potential gives a runaway behaviour to leading order with respect to $\tilde{\rho}.$

In order to avoid any further implicit dependence on $\tau_1,$ we have to choose the implicit dependence in a way that both terms, the D-term contribution and leading order F-term contribution from $\partial W\partial \bar{K} K^{-1} e^K$ have the same $\tau_1$  suppression. Fortunately this can be achieved by the following implicit dependence:
\begin{equation}
|\rho|^2=\frac{|\tilde{\rho}|^2}{\tau_1^{5/6}}=\frac{|\tilde{\rho}|^2}{{\cal V}^{5/9}}.
\end{equation}
At leading order we can stabilise with respect to $\tilde{\rho}$ as follows: Neglecting higher order corrections in $\Phi$, we face the following leading order potential in the limit $a\tau_2\gg 1$ evaluated at $\tilde{\Phi}=\tilde{\Phi}_{\min}:$
\begin{eqnarray}
\nonumber V&=& V_D+V_F\\
\nonumber&=& \frac{2\tau_2^{2n-1} |\tilde{\rho}|^4}{ g_s^{1-2n}{\cal V}^{22/9}}
 +\frac{3M_P^4 g_s\alpha^2 e^{-2a\tau_2}}{2 a \tau_2^{2n-1/2}{\cal V}^{15/9}} \\&+&\frac{2\sqrt{6}g_s^{1/2+n/2}M_P^3 W_0\alpha\sqrt{6}\tau_2^{9/4-5n/6}e^{-a\tau_2}\text{ Re }\tilde{\rho}}{{\cal V}^{22/9}}+\frac{3 g_s M_P^4 W_0^2}{2 g_s^{3/2} {\cal V}^{27/9}}\left(\xi+\frac{g_s^{3/2}n\sqrt{\tau_2}}{a}\right).
 \label{pot1}
\end{eqnarray}
The stabilisation with respect to $\tilde{\rho}$ goes as follows: Assuming $W_0$ to be real, we first observe that all coefficients are in front of $\tilde{\rho}$ are real. Differentiating with respect to $\tilde{\rho}$  and its complex conjugate then directly leads to the fact that $\tilde{\rho}$ has to be real. We can now straightforwardly stabilise $\tilde{\rho}$
\begin{equation}
0=\frac{\partial}{\partial\tilde{\rho}} A (\tilde{\rho})^2(\tilde{\rho}^*)^2+B(\tilde{\rho}+\tilde{\rho}^*)\, ,
\end{equation}
where $A$ and $B$ are abbreviations for the coefficients in the leading order K\"ahler potential. That constraint implies
\begin{equation}
0=2 A \tilde{\rho} (\tilde{\rho}^*)^2+B\overset{\tilde{\rho}\in\mathbb{R}}{=}2 A \tilde{\rho}^3+B\, .
\end{equation}
This can be solved directly for $\tilde{\rho}$ and we obtain in the limit $a\tau_2\gg 1:$
\begin{equation}
 \tilde{\rho}_{\rm min}=\left(\frac{3 a}{2}\right)^{1/6} M_P g_s^{1/2-n/2}(-W_0\alpha)^{1/3} \tau_2^{3/4-5n/6}e^{-a\tau_2/3}.
\end{equation}
After stabilising $\tilde{\rho},$ we end up schematically with the following leading order potential (where we fixed  $n=1/3$ for simplicity):
\begin{eqnarray}
V&= &g_s M_P^4\left(\frac{3\alpha^2 e^{-2a\tau_2}}{2 a \tau_2^{1/6}{\cal V}^{15/9}}+\frac{2 \left(g_s^{1/3}|\tilde{\rho}|^4+\sqrt{6a} M_P^3 W_0 \alpha\text{ Re}(\tilde{\rho})\tau_2^{17/12}e^{-a\tau_2}\right)}{g_s^{5/3}M_P^4\tau_2^{1/3}{\cal V}^{22/9}}+\frac{3 W_0^2\xi}{2 g_s^{3/2}{\cal V}^{27/9}}\right)\nonumber\\
&=& g_s M_P^4\left(\frac{A e^{-2a\tau_2}}{\tau_2^{1/6}{\cal V}^{15/9}}-\frac{B \tau_2^{14/9} e^{-4/3 a\tau_2}}{{\cal V}^{22/9}}+\frac{C}{{\cal V}^{27/9}}\right)\, .
\label{abc}\end{eqnarray}
Where we have defined:
\be
A=\frac{3 \alpha^2}{2 a},\, \qquad B=6^{2/3} \left(1-\left(\frac{1}{2}\right)^{1/3}\right)a^{2/3}(-W_0 \alpha)^{4/3},\qquad  \, C=\frac{3 W_0^2 \xi}{2 g_s^{3/2}}\, .\\
 \ee
Due to the minimisation with respect to $\rho$ we obtain a different exponential suppression than the standard LARGE volume scenario,
but  keep a similar  structure with similar results.

We can clearly see from the potential that the extremisation with respect
to $\tilde{\rho}$ gives a minimum since the coefficients $A$ and $B$ are positive
 and at the minimum we have a negative contribution from this factor.
\subsection{Stabilising the K\"ahler moduli}

To compare our stabilisation procedure to the LARGE volume scenario we go one step back to the potential before extremising with respect to $\tilde{\rho}:$
\begin{eqnarray}
\label{dspot2}
\nonumber V&=& \frac{2\tau_2^{2n-1} |\tilde{\rho}|^4}{ g_s^{1/3}{\cal V}^{22/9}}+\frac{3g_s M_P^4\alpha^2 e^{-2a\tau_2}}{2 a \tau_2^{2n-1/2}{\cal V}^{15/9}}\\
&+&\frac{2\sqrt{6}M_P^3 g_s^{2/3}W_0\alpha\sqrt{6}\tau_2^{5/4-n/2}e^{-a\tau_2}\text{ Re }\tilde{\rho}}{{\cal V}^{22/9}}+\frac{3 M_P^4 W_0^2}{2 {\cal V}^{27/9}}
\left(\frac{\xi}{g_s^{3/2}}+\frac{n\sqrt{\tau_2}}{a}\right).
\end{eqnarray}
This leading order structure looks very promising since it is very close to the original LARGE volume scenario with an additional uplifting term:
\begin{equation}
V_{\text{LV}}=g_s M_P^4\left(\frac{a^2 A^2\sqrt{\tau_2}e^{-2a\tau_2}}{{\cal V}}+\frac{W_0 A a \tau_2 e^{-a\tau_2}}{{\cal V}^2}+\frac{W_0^2 \xi}{{\cal V}^3}+
\frac{V_{\text{uplift}}}{{\cal V}^2}\right) ,
\end{equation}
where we neglected numerical constants for clarity. The different suppressions with respect to the volume are reflected in an overall rescaling. Nevertheless,
 there is a slight difference in the power dependence with regard to the volume in the $\partial W\partial \bar{K} K^{-1} e^K$ ($21/9$ would be perfect) which spoils
the possibility of an exact analytical discussion of the minimisation as it was possible in the LARGE volume case. However the difference to the original setup is
 rather small. Therefore the suppression looks to be a bit larger driving the minimum to larger values. Obviously the adhoc uplifting contribution from previous LARGE
 volume constructions gets replaced by a concrete supersymmetric D-term contribution.\\

After stabilising the matter fields, we now have a potential which only depends on the K\"ahler moduli. Since we are not able to calculate the vacuum structure analytically
 we  find the minima numerically. To be more explicit we will specialise to the modulus weight $n=1/3$ from now on, which is for practical reasons only since similar results will hold for other weights.
 After stabilising the matter fields we obtained the  leading order potential in the limit $a\tau_2\gg 1$ given by (\ref{abc}), which can be written in terms of $\tau_{1,2}$ as:
\begin{equation}
V=g_s M_P^4\left(\frac{A e^{-2 a\tau_2}}{\tau_1^{5/2} \tau_2^{1/6}}-\frac{B e^{-4/3 a \tau_2}\tau_2^{14/9}}{\tau_1^{11/3}}+\frac{C}{\tau_1^{9/2}}\right)
\label{dspot}
\end{equation}
After fixing the matter fields to their value at the minimum, the uplifting term from the D-term seems to disappear since we only have three terms in the potential
 which are similar to the original LARGE volume. However the effect of the D-term potential can be seen in the change of the exponential suppression of the term
 proportional to $B$ which corresponds to the negative contribution. This means that we have a smaller negative contribution compared to the setup without the D-term.
 Despite the fact that the D-term does not seem to be present anymore, it affects the shape of the potential in exactly the way an uplifting term does.

In general, the shape of the potential after fixing the matter fields
suggests that there is a LARGE volume like structure with
  possible D-term lifting. Working strictly within the K\"ahler cone
(i.e. $\tau_2\ll \tau_1$), at relatively small values of the volume
  the term proportional to the $\alpha'$ corrections is dominant.
In an intermediate range the attenuated negative contribution proportional
to $W_0$ can be dominant and at large volumes the leading order
contribution
  will come from the $\partial W\partial \bar{W} K^{-1}$ part. Due to the
domination of this at large volumes, the
potential
  approaches zero from above in all directions. It is in fact
straigtforward to see that when the volume tends to infinity, the
negative term cannot dominate.

  This behaviour at large volume is in agreement with the uplifted LARGE
volume scenario, where we approach zero from above. Therefore, by changing
the
value of the parameters we may obtain an AdS,
almost flat or dS minimum or have a runaway behaviour.
  This  is similar to the lifted LARGE volume scenario but
different from the unlifted one where zero is approached from below, so
we can see the indirect effect of the D-term to uplift the minimum.
Notice also that it is different from the previous ways of getting de
Sitter space from a purely supersymmetric
potential \cite{Balasubramanian:2004uy}, \cite{Westphal:2006tn},
\cite{AbdusSalam:2007pm} in which the stabilised value of the volume was
relatively small, whereas here we are obtaining exponentially large
volumes.\\

Let us study an illustrative example of these possible vacuum structures: First of all, we are indeed able to stabilise the K\"ahler moduli at large values for the large K\"ahler modulus $\tau_1.$ For instance as depicted in
 figure \ref{ads}, we are able to stabilise the K\"ahler moduli with the following naturally chosen parameters:
  \begin{center}
\begin{minipage}{0.7\textwidth}
\centering
\includegraphics[width=0.8\textwidth]{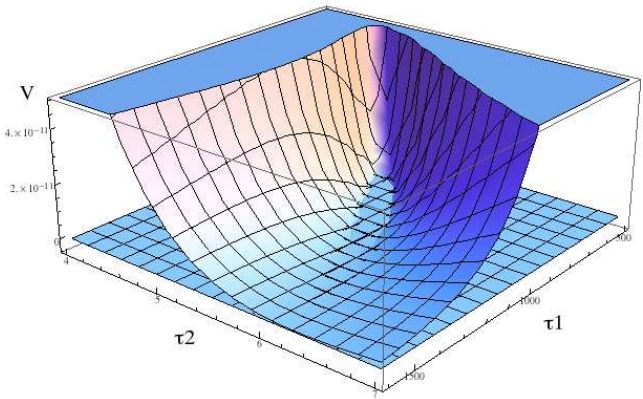}
\end{minipage}\hfill
\begin{minipage}{0.3\textwidth}
\captionof{figure}{\footnotesize{A plot of the numerical
        behaviour of the scalar potential with respect to the K\"ahler moduli showing an AdS minimum.}\label{ads}}
\end{minipage}
\end{center}
\begin{center}

\begin{footnotesize}
\begin{tabular}[h]{c|c|c|c|c|c|c|c|c|c}
$\tau_1$    & $\tau_2$  & $V_{\rm min}$ & $\Phi$    & $\rho$    & $W_0$ & $a$   & $g_s$ & $\alpha$  & $\xi$ \\ \hline
$768.03$ & $5.20$ & $-3.03\times 10^{-12}g_s M_P^4$ & $0.34 g_s^{1/3} M_P$& $0.12 g_s^{1/3}M_P$& $-15.85$& $0.6$& $0.1$& $1.10$& $0.32$
\end{tabular}
\captionof{table}{Showing the parameter values in the AdS-example.}\label{data1}
\end{footnotesize}

\end{center}
In addition to the usual stabilisation corresponding to an AdS geometry, we now can achieve a stabilisation of the K\"ahler moduli that correspond to a dS geometry.
 Taking the previous example, this is achieved by tuning the flux parameter from $W_0=-15.85$ to $W_0=-15.25.$ The resulting figure is \ref{ds} and the numerical values
 in this example are summarised in the following table:
   \begin{center}
\begin{minipage}{0.7\textwidth}
\centering
\includegraphics[width=0.9\textwidth]{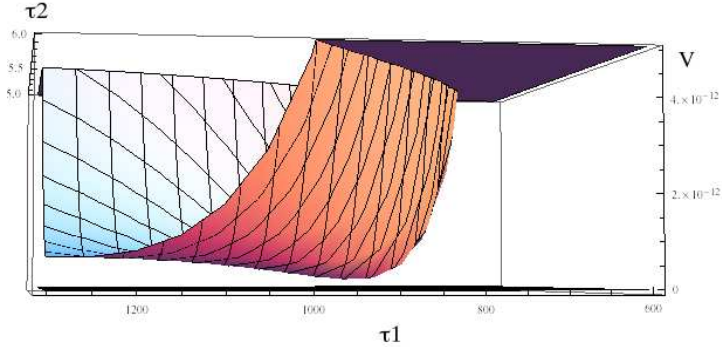}
\end{minipage}\hfill
\begin{minipage}{0.3\textwidth}
\captionof{figure}{\footnotesize{A plot of the numerical
        behaviour of the scalar potential with respect to the K\"ahler moduli showing a dS minimum.}\label{ds}}
\end{minipage}
\end{center}
\begin{center}

\begin{footnotesize}
\begin{tabular}[h]{c|c|c|c|c|c|c|c|c|c}
$\tau_1$    & $\tau_2$  & $V_{\rm min}$ & $\Phi$    & $\rho$    & $W_0$ & $a$   & $g_s$ & $\alpha$  & $\xi$ \\ \hline
$873.40$ & $5.44$ & $1.83\times 10^{-13}g_s M_P^4$ & $0.33 g_s^{1/3} M_P$& $0.11 g_s^{1/3}M_P$& $-15.25$& $0.6$& $0.1$& $1.10$& $0.32$
\end{tabular}
\captionof{table}{Showing the parameter values in the dS-example.}\label{data2}
\end{footnotesize}

\end{center}
Changing the parameters too drastically now results in a disappearance of the minimum as depicted in figure \ref{nom}.
   \begin{center}
\begin{minipage}{0.7\textwidth}
\centering
\includegraphics[width=0.9\textwidth]{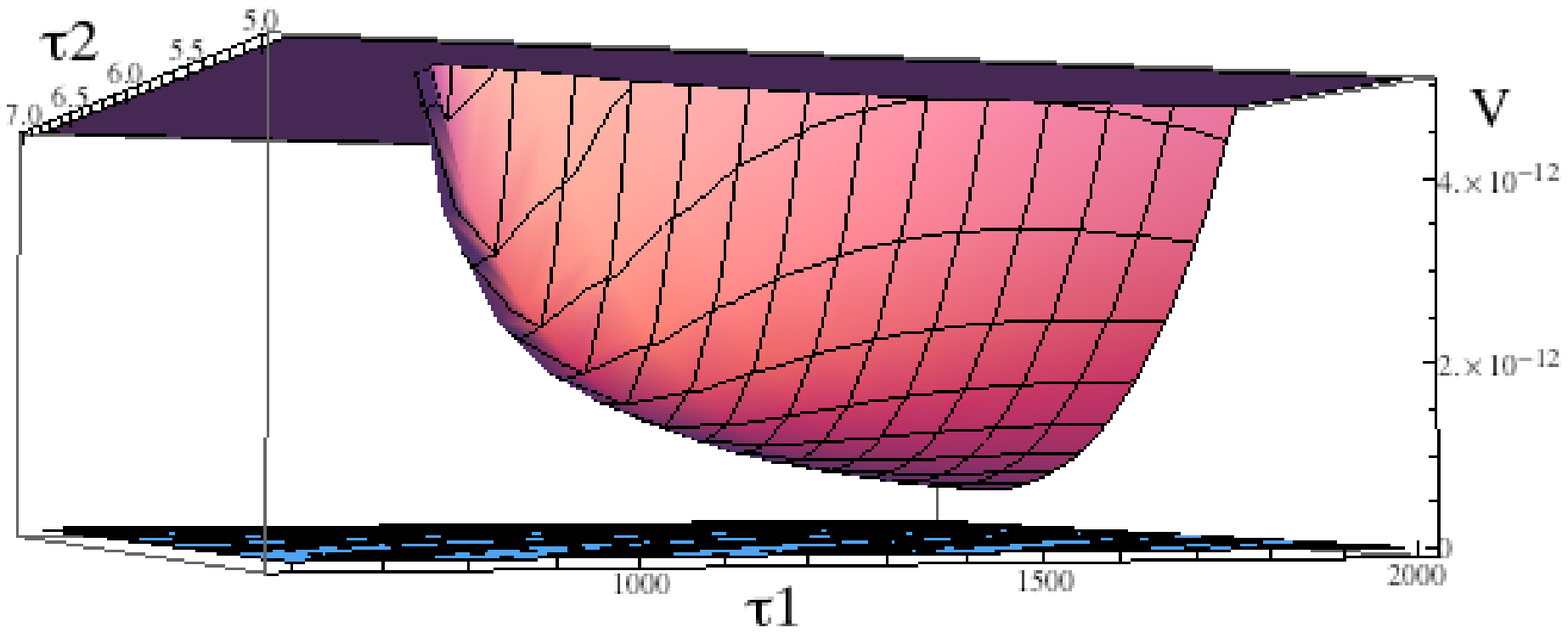}
\end{minipage}\hfill
\begin{minipage}{0.3\textwidth}
\captionof{figure}{\footnotesize{A plot of the numerical
        behaviour of the scalar potential with respect to the K\"ahler moduli showing no minimum.}\label{nom}}
\end{minipage}
\end{center}
In general we find a smooth transition from AdS to dS solutions. We may tune the parameters to obtain almost Minkowski minima \footnote{In principle further tuning freedom can be achieved if the magnetised D7 brane is considered to be in a warped region {\cite{Burgess:2003ic}}    as in the original proposal of \cite{Kachru:2003aw} for the lifting anti-D3 brane.}.
The following example \ref{min} shows a considerable tuning to a minimal value for the potential of $V=5.74\times 10^{-26}.$
   \begin{center}
\begin{minipage}{0.7\textwidth}
\centering
\includegraphics[width=0.9\textwidth]{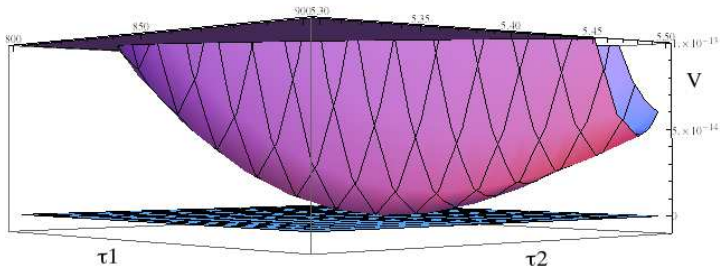}
\end{minipage}\hfill
\begin{minipage}{0.3\textwidth}
\captionof{figure}{\footnotesize{A plot of the numerical
        behaviour of the scalar potential with respect to the K\"ahler moduli with a minimum of $V=5.74\times 10^{-26}.$}\label{min}}
\end{minipage}
\end{center}
For a full analysis we should check our analytical results for the matter fields by visualising the potential with respect to the matter fields as well.
In the case of the dS example from the previous section we obtain the following figures \ref{matterrho} and \ref{matterphi}.
     \begin{center}
\begin{minipage}{0.48\textwidth}
\centering
\includegraphics[width=0.95\textwidth]{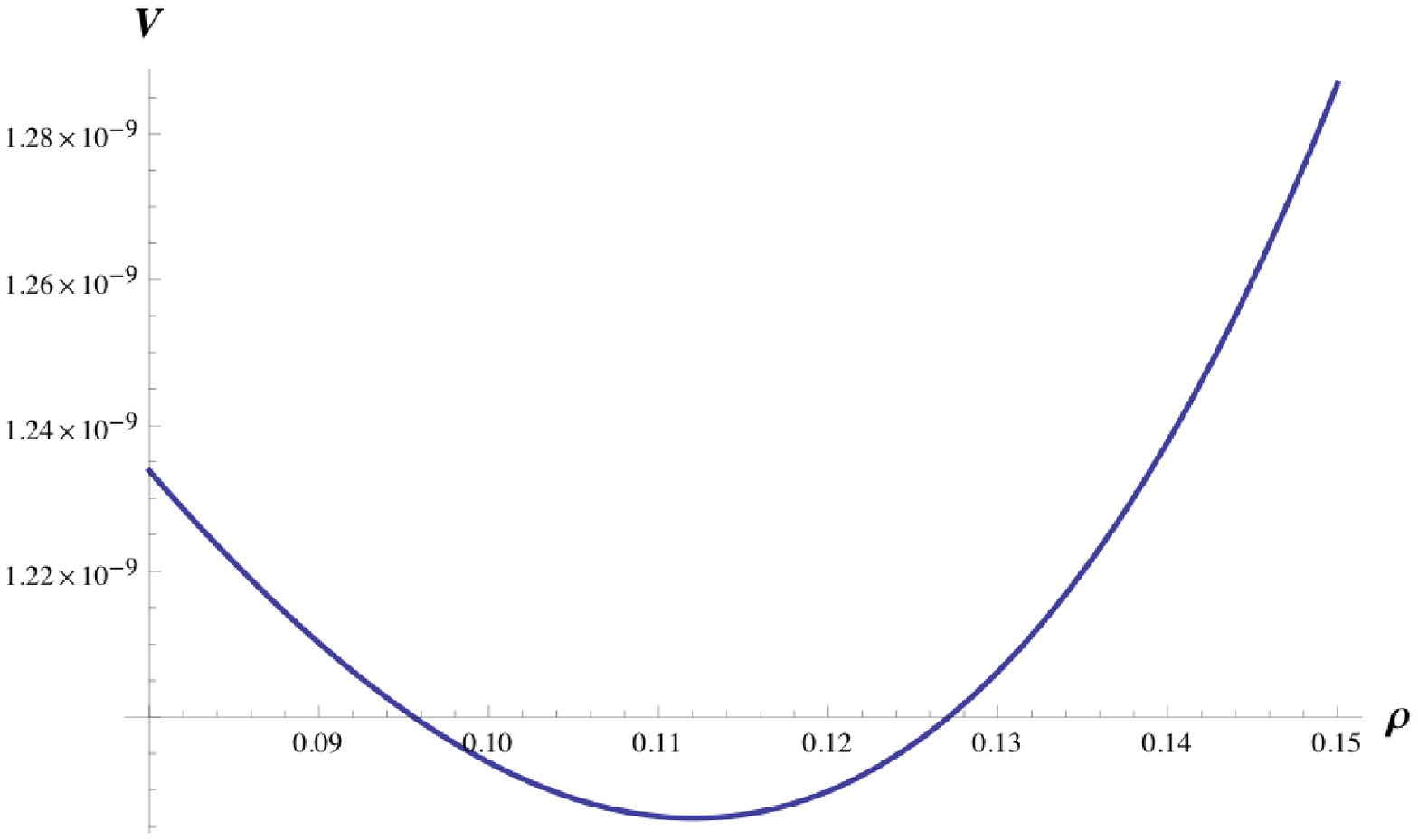}
\captionof{figure}{\footnotesize{A plot of the numerical
        behaviour of the scalar potential with respect to the matter field $\rho$.}
    \label{matterrho}}
\end{minipage}\hfill
\begin{minipage}{0.48\textwidth}
\includegraphics[width=0.95\textwidth]{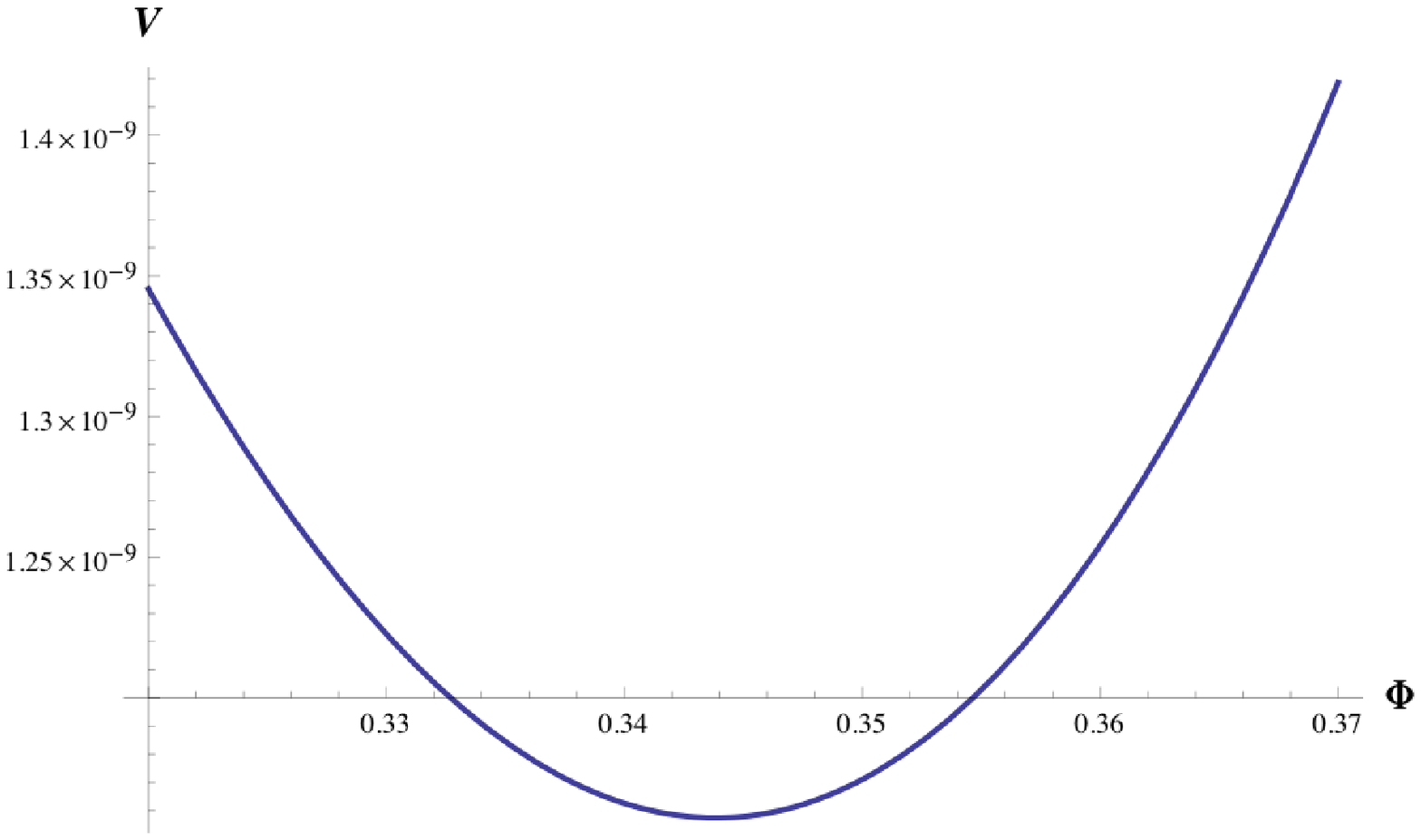}\\
\captionof{figure}{\footnotesize{A plot of the numerical
        behaviour of the scalar potential with respect to the matter field $\Phi$.}
    \label{matterphi}}
\end{minipage}
\end{center}
Both plots support our previous analysis with a minimum in the predicted region. In particular the correction to $\Phi$ does not spoil our uplifting mechanism.
\section{D7 Soft-Terms}
We now would like to study the mechanism of supersymmetry breaking within the 2-moduli framework presented in the previous sections. For this we assume a brane construction on a separate four cycle $\tau_3$ (e.g. through an $SU(5)$ GUT model \cite{Blumenhagen:2008zz}) leading at low-energies to a MSSM realisation. The Standard Model cycle $\tau_3$ cannot be stabilised by non-perturbative effects as pointed out in \cite{Blumenhagen:2007sm}, however loop effects could stabilise the additional cycle \cite{Cicoli:2008va}, leaving us for now with a regime of effective field theory ($\tau_3\neq 0$). This ansatz is slightly different from the original study of soft-terms in the LARGE volume scenario (cf. \cite{Conlon:2005ki} and \cite{Conlon:2006wz}), but was recently discussed in \cite{Blumenhagen:2009gk}. \footnote{We concentrate here on generic D7 brane soft terms without including explictly the mechanism for stabilising the MSSM modulus. We only assume that it is not stabilised at a singular point. A detailed study of this consideration is left for future work.}.

The volume in this extended Swiss-cheese Calabi-Yau 
can be written as ${\cal V}=\tau_1^{3/2}-\tau_2^{3/2}-\tau_3^{3/2}.$ The philosophy of our approach is depicted in figure \ref{swisscheese} below.
   \begin{center}
\begin{minipage}{0.65\textwidth}
\centering
\includegraphics[width=0.8\textwidth]{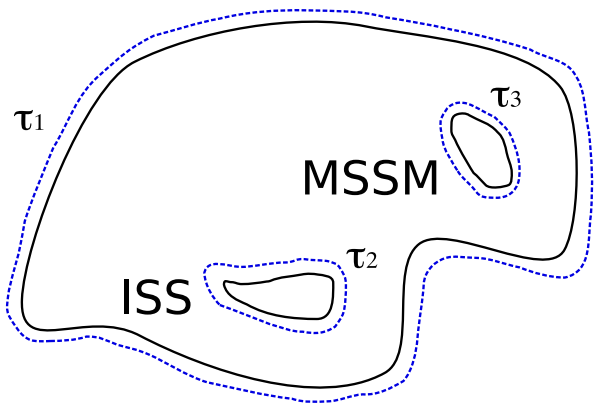}
\end{minipage}\hfill
\begin{minipage}{0.35\textwidth}
\captionof{figure}{\footnotesize{In the simplest model the geometry of the Calabi-Yau is parametrised by three 4 cycles. $\tau_1$ is the large cycle; $\tau_2,$ $\tau_3$ denote the small (blow-up) cycles. We assume the MSSM to be localised at $\tau_3$ and our ISS brane setup is localised on $\tau_2.$}\label{swisscheese}}
\end{minipage}
\end{center}
We start with the following general effective supergravity ansatz:
\begin{eqnarray}
K&=&-2M_P^2 \log{\left({\cal V}+\xi s^{3/2}\right)}-\log{(s)}+\frac{\tau_2^n}{\tau_1}\left(|p|^2+|q|^2+|\rho|^2+|\Phi|^2\right)+\tilde{K}^i |\upsilon_i|^2\, ,\\
W&=& M_P^3 g_s^{3/2} W_0+ M_P g_s^{1/2+n} \alpha e^{-a T_2}\left(\frac{p\Phi q}{\mu}+\rho\Phi\right)+W_{\rm MSSM}\, ,
\end{eqnarray}
where we reintroduced the dilaton dependence $s;$ $\upsilon$ denotes one of the chiral superfields within the MSSM whose superpotential is given by $W_{\rm MSSM}.$ For now, the correct form of the moduli weights in the K\"ahler potential is not of major importance, and we keep our study as general as possible.

A priori we can say that the scheme of supersymmetry breaking is gravity (moduli) mediated. The amount of gauge mediated contribution can be seen in the fact that the masses associated with the hidden sector matter fields $\rho$ and $\Phi$ are larger or of the order of the gravitino mass (see the appendix for an estimate of the masses associated to our model). In due course we will establish the same or even stronger contributions to the moduli mediated 
soft masses compared to the LVS and we therefore can neglect contributions arising from anomaly mediation.

We follow the standard mechanism of calculating soft supersymmetry breaking terms as for example used and described in \cite{Conlon:2006gv}.

\subsection*{Gravitino and Gaugino Masses}
The gravitino mass is found to have the standard volume suppression:
\begin{eqnarray}
\nonumber m_{3/2}&=&e^{\frac{K}{2 M_P^2}}\frac{|W|}{M_P^2}\, \\
&=& \frac{M_P |W_0| g_s^{1/2}}{{\cal V}_E}\, .
\end{eqnarray}
where the subscript $E$ is added to clarify that this is the Einstein-frame volume.

The F-terms needed for the remaining soft-terms are calculated in detail in appendix F. Gaugino masses are in general given by
\begin{equation}
M_a=\frac{1}{2} ({\rm Re }\, f_a)^{-1} F^m \partial_m f_a\, .
\end{equation}
Assuming our standard model is constructed via D7 branes, the gauge kinetic function is
\begin{equation}
f_{D7}=T_3\, ,
\end{equation}
 Using those results, we find for the D7 brane gaugino masses the following mass:
\begin{eqnarray}
M_{D7}&\sim&\frac{W_0 M_P g_s^{1/2}}{2 {\cal V}_E} =\frac{m_{3/2}}{2}\, .
\end{eqnarray}
Regarding the volume suppression of the gaugino masses, the results coincide with the LARGE volume analysis. The additional suppression of the gaugino masses due to a next to leading order cancellation in the F-terms (discussed in \cite{Conlon:2006us}) does not occur since $\tau_3$ is not stabilised by non-perturbative effects.
\subsection*{Scalar Masses}
Scalar masses can be found by the following formula
\begin{equation}
m_i^2=m_{3/2}^2-F^m \bar{F}^{\bar{n}}\partial_m\partial_{\bar{n}}\log{\tilde{{\cal K}}_i}\, ,
\end{equation}
where we neglect contributions from the vacuum energy.\footnote{Here the addition of D-terms does not change the formula for scalar masses since the SM fields are not charged under the anomalous $U(1)$ in the hidden sector. A general formula was presented for example in \cite{Dudas:2005pr}.} To calculate the scalar masses we have to specify the matter metric $\tilde{{\cal K}}_i.$ However we can start with the following general ansatz
\begin{equation}
\tilde{{\cal K}}_i\sim \frac{\tau_3^b s^c}{\tau_1^d}\, .
\end{equation}
This gives the following leading order results
\begin{eqnarray}
\partial_{\tau_1}\partial_{\tau_1}\log{\tilde{{\cal K}}_i}&=&\frac{d}{4 \tau_1^2}\, ,\\
\partial_{\tau_3}\partial_{\tau_3}\log{\tilde{{\cal K}}_i}&=&-\frac{b}{4 \tau_3^2}\, ,\\
\partial_{s}\partial_{s}\log{\tilde{{\cal K}}_i}&=&-\frac{c}{4 s^2}\, .
\end{eqnarray}
Using the results for the F-terms derived in the appendix, we find the following results:
\begin{eqnarray}
(F^{\tau_1})^2\partial_{\tau_1}\partial_{\tau_1}\log{\tilde{{\cal K}}_i}&\sim& \frac{ d W_0^2 M_P^2 g_s}{{\cal V}^2}\\
(F^{\tau_3})^2\partial_{\tau_3}\partial_{\tau_3}\log{\tilde{{\cal K}}_i}&\sim& -\frac{ b M_P^2 W_0^2 g_s}{{\cal V}^2}\\
(F^{s})^2\partial_{s}\partial_{s}\log{\tilde{{\cal K}}_i}&\sim& - \frac{c M_P^2 W_0^2 g_s}{{\cal V}^4}
\end{eqnarray}
We can directly see that the contribution arising from the dilaton
is negligibly small. In addition, we find that the leading order
term proportional to the gravitino mass vanishes if $d-b=1.$ If this
relation is satisfied, we have to identify the next to leading order
contribution. Unlike in the original LARGE volume analysis, there is no sub-leading cancellation in the F-terms \cite{Blumenhagen:2009gk} for the K\"ahler moduli since the $e^{-2a\tau_2}$ term in (\ref{dspot}) comes from $\partial_{\phi}W\partial_{\phi} W K^{\phi\phi}$ rather than $\partial_{\tau_2}W\partial_{\tau_2} W K^{\tau_2\tau_2}$ and the D-term uplifting contribution require a slightly different stabilisation as discussed in the previous section. This determines the sub-leading contribution (see appendix for further details) to come from the mixing of the non-perturbative leading
order contribution to the F-term and the other leading order F-term
contribution. We obtain the interesting result:
\begin{eqnarray}
\label{nonpet}
F^{\tau_1} F^{\tau_1}_{\rm np}\partial_{\tau_1}\partial_{\tau_1}\log{\tilde{{\cal K}}_i}&\sim& \frac{2 d W_0 g_s^{m} \tilde{\rho}\tilde{\Phi}\tau_2 e^{-a \tau_2}}{{\cal V}^2 \tau_1^{2/3}}\, ,\\
F^{\tau_3} F^{\tau_3}_{\rm np}\partial_{\tau_3}\partial_{\tau_3}\log{\tilde{{\cal K}}_i}&\sim& -\frac{2 b W_0 g_s^{m} \tilde{\rho}\tilde{\Phi}\tau_2 e^{-a \tau_2}}{{\cal V}^2 \tau_1^{2/3}}\, .
\end{eqnarray}
Evaluating these contributions at the LARGE volume minimum ($e^{a\tau_2}\sim {\cal V}^x$) gives the following common volume suppression:
\begin{equation}
F^{\tau_i} F^{\tau_i}_{\rm np}\partial_{\tau_i}\partial_{\tau_i}\log{\tilde{{\cal K}}_i}\sim \frac{d-b}{{\cal V}^{2+\frac{4}{9}+\frac{4x}{3}}}\, .
\end{equation}
Overall we find the general expression for the scalar masses to be
\be
m_i^2=(1-d+b)m_{3/2}^2-2 (d-b)\frac{W_0 g_s^{m} \tilde{\rho}\tilde{\Phi}\tau_2 e^{-a \tau_2}}{{\cal V}^2 \tau_1^{2/3}}\, ,
\ee
where the second term is subleading since it is higher suppressed with respect to the volume.

For example, taking $d=1$ and $b=1/3$ as in previous sections and in the analysis of soft terms in the LVS \cite{Conlon:2006wz}, we obtain no cancellation at leading order and find the following scalar masses:
\begin{equation}
m_i^2=\frac{1}{3} m_{3/2}^2+{\rm higher\, order\, corrections.}
\end{equation}
\subsection*{A-terms}
The A-terms are given by
\begin{equation}
A_{ijk}=F^m(\partial_m K+\partial_m \log{Y_{ijk}}-\partial_m\log{\tilde{K}_i\tilde{K}_j\tilde{K}_k})
\end{equation}
Assuming a constant $Y_{ijk}$ we can estimate the A-terms to be given by the following expression at leading order:
\begin{eqnarray}
\nonumber A_{iii}&=&-\frac{2 \tau_1 \sqrt{g_s} |W_0| M_P}{{\cal V}}\left(\frac{3}{2\tau_1}-\frac{3 d}{2\tau_1}\right)-\frac{2 \tau_3 \sqrt{g_s} |W_0| M_P}{{\cal V}}\left(\frac{3 \sqrt{\tau_3}}{2{\cal V}}+\frac{3 b}{2\tau_3}\right)+\frac{2 \tau_2 \sqrt{g_s} |W_0| M_P}{{\cal V}}\left(\frac{3 \sqrt{\tau_2}}{2{\cal V}}\right)\\
&=&- 3 m_{3/2} (1-d+b)+\text{ h. o.}\, ,
\end{eqnarray}
which are universal and of the order of the gravitino mass.
\subsection*{Short Summary}
Overall we find three different scenarios according to the value of
$d-b$ and $\tau_3$:
\begin{enumerate}
\item If $d-b\neq 1$ the soft terms are of the order the gravitino
mass.
\item If $d-b=1$ the gaugino masses are of order of the gravitino
mass but scalar masses and $A$-terms are much more suppressed.
\end{enumerate}
In all cases the soft terms are universal \cite{Conlon:2006wz,
Conlon:2007dw}.

We can compare these results with the original LARGE volume analysis
from \cite{Conlon:2006wz} where cancelations in the F-terms occurred
due to the minimization of the small cycles via non-perturbative
effects and the soft-terms where found to be: 
$$M_i\sim \frac{m_{3/2}}{\log m_{3/2}}, \,  m_i=  n M_i, \,
A_{\alpha\beta\gamma} =  -3n M_i\, .$$

 The additional
small suppression of the soft-terms compared to the gravitino mass found
in the original LARGE volume analysis is not present here and this
feature distinguishes the structure of soft-terms from both
scenarios.

The difference to the corresponding 3-cycle set-up in the LVS \cite{Blumenhagen:2009gk} is that the slight difference in the potential (D-term uplifting contribution especially) renders the scalar masses not affected by sub-leading corrections in the F-term or the K\"ahler metric. A detailed quantitative analysis of this scenario is out of the scope of this article.
\section{K\"ahler moduli inflation}
In the previous sections we described a mechanism that allows us to obtain dS vacua within a fully ${\cal N}=1$ supersymmetric action. We now would like to show how to implement the model of K\"ahler moduli inflation \cite{Conlon:2005jm} within this framework.

The starting point of the discussion is the modification of the original geometry of $\mathbb{P}_{[1,1,1,6,9]}$ by adding an additional small 4-cycle. The volume is modified to
\begin{equation}
{\cal V}=\tau_1^{3/2}-\tau_2^{3/2}-\tau_3^{3/2}\,.
\end{equation}
On that additional 4-cycle we assume a non-perturbative effect $B e^{-bT_3}.$ The K\"ahler and superpotential are changed to:
\begin{eqnarray}
K&=&-2M_P^2\log{\left({\cal V}+\frac{\xi}{g_s^{3/2}}\right)}+\frac{g_s^n\tau_2^{n}}{g_s\tau_1}(|p|^2+|q|^2+|\rho|^2+|\Phi|^2)\, ,\\
W&=&M_P^3 g_s^{3/2}W_0+M_P g_s^{1/2+n}e^{-a T_2} \alpha(\frac{p\Phi q}{\mu}+\Phi \rho)+M_P^3 g_s^{3/2} B e^{-b T_3}\, .
\end{eqnarray}
This change alters the leading order potential to:
\begin{equation}
V=V_{\rm old}+g_s M_P^4\left(\frac{8 (b B)^2 \sqrt{\tau_3}e^{-2b \tau_3}}{3{\cal V}}+\frac{4W_0 b B\tau_3 e^{-b\tau_3}}{{\cal V}^2}\right),
\end{equation}
where $V_{\rm old}$ is the potential discussed in previous sections. The old potential is independent of $\tau_3$ and we can hence stabilise the new part with respect to $\tau_3$ at constant volume. We obtain in the limit $bT_3\gg 1:$
\begin{equation}
bB e^{-bT_3}=\frac{W_0 \sqrt{\tau_3}}{12 {\cal V}}.
\end{equation}
Plugging this value into the original potential gives the following contribution at the extremal value for $\tau_3:$
\begin{equation}
V=V_{\rm old}-\frac{17 W_0^2 (\tau_3^{\rm 0})^{3/2}}{54 {\cal V}^3}\, .
\end{equation}
Since the contribution from the potential including $\tau_3$ is negative, we have to be at a minimum with respect to $\tau_3.$ Having a negative contribution from the inflationary potential at the minimum then also requires to stabilize the potential at a ``large'' positive value and not at an almost Minkowski minimum. The assumption that ${\cal V}$ can be taken to be constant during inflation is justified in the limit $b\gg a$ as the effect of the additional cycle $\tau_3$ towards the potential is negligibly small compared to the old potential. Hence we end up with the same potential as in the K\"ahler moduli inflation but with the advantage that we can use our new mechanism to stabilise the volume at a positive vacuum energy.

Following the analysis of the original paper \cite{Conlon:2005jm}, we can now calculate the inflationary characteristics of this model. Our inflationary potential is given by
\begin{equation}
V=V_{\rm old}+\frac{4 \tau_3 W_0 b B e^{-b \tau_3}}{{\cal V}^2}\, ,
\end{equation}
where we neglect the higher suppressed and hence irrelevant term including the double exponential. The physically relevant parameter is the canonically normalised version of $\tau_3$ which is found to be at leading order
\begin{equation}
\tau_3^c=\sqrt{\frac{3}{4}}\frac{\tau_3 M_P}{(\tau_1^0)^{3/4}(\tau_3^0)^{1/4}},
\end{equation}
where the exponent "$0$" denotes that the value is taken at the minimum. Rewriting the potential in terms of the canonically normalised field $\tau_3^c$ shows the exponential suppression with respect to the volume
\begin{equation}
V=V_0+\frac{4 W_0 b B}{M_P{\cal V}^2}\left(\frac{3 {\cal V}}{4}\right)^{2/3}(\tau_3^c)^{4/3} \exp{\left[-b \left(\frac{3 {\cal V}}{4}\right)^{2/3}\frac{(\tau_3^c)^{4/3}}{M_P}\right]}.
\end{equation}
From this potential we find the following slow-roll parameters:
\begin{eqnarray}
\epsilon&=&\frac{M_P^2}{2}\left(\frac{V'}{V}\right)^2=\frac{32 (W_0bB)^2}{3 V_0^2 {\cal V}^4}\tau_3^{1/2}(1-b\tau_3)^2 e^{-2b\tau_3}\, ,\\
\eta&=&M_P^2 \frac{V''}{V}=\frac{4 W_0 bB}{3 \sqrt{\tau_3} V_0 {\cal V}^2}(1-9b \tau_3+4(b\tau_3)^2)e^{-b\tau_3}\, ,\\
\xi&=&M_P^4 \frac{V'V'''}{V^2}=- \frac{32 (W_0 b B)^2}{9 V_0^2{\cal V}^4\tau_3}(1-b\tau_3)(1+11b\tau_3-30 (b\tau_3)^2+8 (b\tau_3)^3)e^{-2b \tau_3}\, ,
\end{eqnarray}
where the derivatives are taken with respect to the canonically normalised fields. In these parameters the only small difference is that we have not specified $V_0$ yet, which will be of a similar form compared to the LARGE volume scenario. From the slow-roll parameters one can determine the spectral index and its running as
\begin{eqnarray}
n-1&=&2\eta-6\epsilon + O(\xi)\, ,\\
\frac{dn}{d \ln k}&=& 16 \epsilon\eta-24 \epsilon^2-2\xi\, .
\end{eqnarray}
In analogy to the original discussion of K\"ahler moduli inflation we find the number of e-foldings to be given by
\begin{equation}
N_e=\int_{\phi_{\rm end}}^{\phi}\frac{V}{V'}d\phi=\frac{3 V_0}{16 W_0 bB}\int_{\tau_3^{\rm end}}^{\tau_3}\frac{e^{b\tau_3}}{\sqrt{\tau_3}(1-b\tau_3)}d\tau_3\, .
\end{equation}
To match the COBE normalisation for the density fluctuations $\delta_H=1.92\times 10^{-5}$ we have to satisfy the following constraint
\begin{equation}
\frac{V^{3/2}}{M_P^3 V'}=5.2\times 10^{-4},
\end{equation}
where the potential is evaluated at the horizon exit, which means $N_e=50-60$ e-foldings before the end of inflation.
This endows us with a constraint to determine the contribution from the original potential $V_0.$ In general the only modification on the model of K\"ahler moduli inflation is given by the fact that the old LARGE volume contribution is replaced by the new potential which was developed and discussed in the previous sections. It is therefore most likely that a concrete calculation, which we have not completed yet, will give the same numerical results as in the original K\"ahler moduli scenario, which are given by:
\begin{itemize}
\item The tensor-to-scalar ratio was found to be
\begin{equation}
r\sim 16\epsilon\, .
\end{equation}
\item For $50-60$ e-foldings, the model gives rise to the following characteristics:
\begin{eqnarray}
0.960 < &n& < 0.967\, ,\\
-0.0006< &\frac{dn}{d\ln k}&< -0.0008\, ,\\
0< &|r|&< 10^{-10}\, ,\\
10^5 l_s^6\leq &{\cal V}&\leq 10^7 l_s^6\, .
\end{eqnarray}
\end{itemize}
\section{3-Parameter K3 fibration and Fibre Inflation}
To show the generality of the uplifting mechanism, we now consider the example of a 3-parameter K3 fibration which allows K\"ahler moduli stabilisation at LARGE volume (cf. \cite{Cicoli:2008va} and \cite{Cicoli:2008gp}). We start with the same expression for the volume as in the mentioned articles
\begin{equation}
{\cal V}=\alpha \sqrt{\tau_1}(\tau_2-\beta \tau_1)-\gamma\tau_3^{3/2}\, .
\end{equation}
It was shown that the combination $\alpha \sqrt{\tau_1}(\tau_2-\beta \tau_1)$ plays the r\^ole of the exponentially dominating volume as in the original LARGE volume scenario. $\tau_3$ plays the role of a blow-up and is crucial for the existence of a stable minimum at LARGE volume. Taking only the scalar potential, one still remains with one flat direction, corresponding essentially to $\tau_1.$ It was shown in \cite{Cicoli:2008va} that this runaway behaviour can be stabilised by considering loop-corrections to the potential.

To embed our uplifting scenario in this K3 setup, we need to know where (i.e. on which cycle) non-perturbative effects are required. To keep the same structure of moduli stabilisation it is necessary to place our brane setup on the blow-up cycle $\tau_3.$ Since both of the other cycles have to be too large and the non-perturbative effects are hence negligibly small on those cycles.\\
 Our setup in terms of the K\"ahler and superpotential then looks like
\begin{eqnarray}
K&=&-2M_P^2\log{\left({\cal V}+\frac{\xi}{g_s^{3/2}}\right)}+\frac{\tau_{3}^m}{g_s^{1-m}{\cal V}^{2/3}}\left(|p|^2+|q|^2+|\rho|^2+|\Phi|^2\right),\\
W&=&g_s^{3/2}M_P^3 W_0+M_Pg_s^{1/2+m}\delta e^{-a T_{3}}\left(\frac{p\Phi q}{\mu}+\rho\Phi\right).
\end{eqnarray}
Wrapping around the 4-cycle parametrised by $\tau_3$ gives us the following leading order structure in the D-term potential:
\begin{eqnarray}
V_D&=&\frac{1}{2\tau_3}\left(\frac{\tau_3^m}{{\cal V}^{2/3}}\left(q_\rho |\rho|^2+q_\Phi |\Phi|^2\right)+Q_{\tau_3}\left(\frac{3 \sqrt{\tau_3}}{2({\cal V}+\xi)}+\frac{m \tau_3^{m-1}(|\rho|^2+|\Phi|^2)}{{\cal V}^{2/3}}\right)\right)^2\, ,
\end{eqnarray}
where we directly set the quark fields to zero. The structure of the D-term potential is essentially the same as in the analysis of the $\mathbb{P}_{[1,1,1,6,9]}$ geometry from the previous section. This not only allows us to set the quark fields to zero but also to cancel the FI-term with the least suppressed $\Phi-$field.\\
Hence obtain the following implicit dependence of $\Phi$ on the large K\"ahler moduli
\begin{equation}
|\Phi|^2=\frac{|\tilde{\Phi}|^2}{{\cal V}^{1/3}}=\frac{Q_{\tau_3}{\cal V}^{2/3} \tau_3^{\frac{1}{2}-m}}{2 q_{\Phi}({\cal V}+\xi)}(1+\frac{m Q_{\tau_3}}{q_\Phi \tau_3})^{-1}\approx\frac{Q_{\tau_3} \tau_3^{\frac{1}{2}-m}}{2q_\Phi {\cal V}^{1/3}}\, .
\end{equation}
As in the Swiss-cheese case we can only assume at this stage that $\rho$ is implicitly higher suppressed than $\Phi$ with respect to the K\"ahler moduli.

With this assumption and the knowledge from the previous study in the Swiss-cheese case we can now estimate the leading order contributions to the F-term potential:
\begin{eqnarray}
e^K&\sim& \frac{1}{{\cal V}^2}\\
\partial W\partial \bar{W}K^{-1}&\sim&g_s^{2+m}M_P^2 \delta^2 e^{-2a\tau_3}\frac{ {\cal V}^{1/3} |\tilde{\Phi}|^2}{\tau_3^m}\\
\partial W\partial \bar{K}K^{-1} \bar{W}+\text{c. c.}&\sim&\frac{12 W_0 g_s^{3/2+m}M_P^3 \delta \text{Re } {\tilde{\rho}\tilde{\Phi}}e^{-a\tau_3}\gamma  {\cal V}^{1/3}\tau_3(m+a\tau_3)}{\left(3 g_s M_P^2 \gamma  {\cal V}^{1/3}\tau_3^{3/2}+2 g_s^m m \alpha \sqrt{\tau_1}(\beta\tau_1-\tau_2)\frac{\tau_3^m|\tilde{\Phi}|^2}{ {\cal V}^{1/3}}\right) {\cal V}^{1/6+n}}\\
\partial K\partial \bar{K}K^{-1}&\sim&3|W_0|^2+W_0^2\frac{A\xi+B|\tilde{\Phi}|^2}{\tau_b^{3/2}}\, ,
\end{eqnarray}
where the coefficients $A$ and $B$ are introduced to keep the overall structure feasible. $\tau_b$ denotes the power suppression with respect to the large K\"ahler moduli.

We can now determine the implicit dependence of $\rho$ on the K\"ahler moduli in the same way as before to be $n=5/18.$ This enables us to integrate out the matter fields completely and we end up with the following potential with respect to the K\"ahler moduli
\begin{eqnarray}
V&=&\frac{A \tau_3^{1/2-2m}e^{-2a\tau_3}}{{\cal V}^{5/3}}-\frac{B e^{-\frac{4}{3}a\tau_3}\tau_3^{5/3-2m}}{ {\cal V}^{\frac{7}{3}+\frac{1}{9}}}+\frac{C}{ {\cal V}^{3}}\, .
\end{eqnarray}
Let us compare this potential with the original 3 parameter K3 potential which was calculated in \cite{Cicoli:2008va} to be given by
\begin{equation}
V=\frac{A\sqrt{\tau_3}e^{-2a\tau_3}}{{\cal V}}-\frac{B \tau_3 e^{-a\tau_3}}{{\cal V}^2}-\frac{C}{{\cal V}^3}\, .
\end{equation}
The comparison leads to exactly the same results as in the $\mathbb{P}_{[1,1,1,6,9]}$ analysis from the previous section:
\begin{itemize}
\item We have a marginally higher suppression with respect to the overall volume arising from the change in the leading order F-terms. In addition we have a slightly larger suppression with respect to the volume in the second term, due to the matching of $\rho$ D-term and F-term contributions with powers of the large K\"ahler moduli.
\item In the same term we have the only other difference in the exponential suppression arising after integrating out $\rho.$
\item Despite missing the analytic minimisation with respect to the volume by this slight change, we can still approximate the minimisation. We then see that we have the typical exponential hierarchy between the volume and the blow-up K\"ahler modulus.
\item To stabilise the remaining large K\"ahler modulus we can still use loop corrections as discussed in \cite{Cicoli:2008va}.
\end{itemize}
Due to the same change in the potential, we assume at this stage that it is possible to stabilise the K\"ahler moduli in the same fashion with the additional uplifting as in the Swiss-cheese case.
\subsection{Fibre Inflation}
Following the successful embedding of D-term uplifting to the 3-Parameter K3 fibration it is natural to raise the question whether we can embed fibre inflation \cite{Cicoli:2008gp} into this uplifting scenario. In order to realise the proposal of fibre inflation we need to wrap branes around the two large cycles $\tau_1$ and $\tau_2.$ These additional branes do not intersect with the blow-up cycle and do not create additional matter field content at the intersections. To obtain the potential of fibre inflation, one has to study the loop corrections ($g_s$) to the scalar potential. As shown in section 3.1.2 of \cite{Cicoli:2008gp} the leading order string-loop corrections are then given by the following contributions:
\begin{itemize}
\item From the branes wrapping the 4-cycle $\tau_1:$
\begin{equation}
\delta V^{\rm KK}_{(g_s),\tau_1}=\frac{g_s^2 W_0^2 (C_1^{\rm KK})^2}{\tau_1^2{\cal V}^2}.
\end{equation}
\item From the branes wrapping the 4-cycle $\tau_2:$
\begin{equation}
\delta V^{\rm KK}_{(g_s),\tau_2}=\frac{2 g_s^2 W_0^2 (C_2^{\rm KK})^2}{\tau_2^2{\cal V}^2}.
\end{equation}
\item From the intersections of the two stacks of branes around $\tau_1$ and $\tau_2:$
\begin{equation}
\delta V^{\rm KK}_{(g_s),\tau_1\tau_2}=-\frac{2 C_{12}^W}{\sqrt{\tau_1}}\frac{W_0^2}{{\cal V}^3}.
\end{equation}
\item From the branes wrapping the blow-up cycle:
\begin{equation}
\delta V^{\rm KK}_{(g_s),\tau_3}=\frac{g_s^2 W_0^2 (C_3^{\rm KK})^2}{\sqrt{\tau_3}{\cal V}^3}.
\end{equation}
This contribution does not depend on the "flat"-direction $\tau_1$ and can be hence understood as a subleading
correction in the $\alpha'$-corrections.
\end{itemize}
With those results, we can establish our additional inflationary potential on top of the leading order scalar potential of the previous section:
\begin{eqnarray}
V&=&\frac{A \tau_3^{1/2-2m}e^{-2a\tau_3}}{{\cal V}^{5/3}}-\frac{B e^{-\frac{4}{3}a\tau_3}\tau_3^{5/3-2m}}{ {\cal V}^{\frac{7}{3}+\frac{1}{9}}}+\frac{C}{ {\cal V}^{3}}+ V_{\rm{inf}}\\
&=&\frac{A \tau_3^{1/2-2m}e^{-2a\tau_3}}{{\cal V}^{5/3}}-\frac{B e^{-\frac{4}{3}a\tau_3}\tau_3^{5/3-2m}}{ {\cal V}^{\frac{7}{3}+\frac{1}{9}}}+\frac{C}{ {\cal V}^{3}}+\frac{W_0^2}{{\cal V}^2}\left(\frac{D}{\tau_1^2}-\frac{E}{{\cal V} \sqrt{\tau_1}}+\frac{F \tau_1}{{\cal V}^2}\right).
\end{eqnarray}
We see that as in the discussion of K\"ahler moduli inflation both parts of the potential simply decouple.
Exactly as in the original discussion of fibre inflation it is possible to stabilise the volume in a first step and then look at the $\tau_1$ direction as the inflationary direction. We can therefore say that it is straight forward to embed fibre inflation into the uplifted scenario and under the assumption of a constant volume the analysis of inflationary parameters will give exactly the same results. Although a multi-field analysis (i.e. taking the volume to be non-constant) is out of the scope of this article, we can still comment on the stability of our uplifted scenario with regard to variations of the volume. It was observed in the analysis of the fibre inflation model that it is necessary to avoid a runaway in the volume to introduce an uplifting term proportional to $1/{\cal V}^{4/3}.$ In our scenario, we exactly have such an uplifting term before integrating out $\rho,$ since, as previously discussed, the uplifting is caused by
\begin{equation}
V_{\rm up}\approx \frac{|\tilde{\rho}|^4}{{\cal V}^{22/9}}=\frac{|\rho|^4}{{\cal V}^{4/3}}\,.
\end{equation}
We therefore expect that the multi-field analysis would give exactly the same results as in the original analysis.
\section{Constraints for metastable dS vacua in supergravity setups}
In recent years it was studied under which general constraints it is possible to obtain metastable dS vacua and/or inflation in a general supergravity framework \cite{GomezReino:2008px,Covi:2008ea,GomezReino:2007qi,Covi:2008cn}. Since our construction from previous sections is presented in a fully ${\cal N}=1$ supergravity framework we would like to comment on why our approach satisfies these constraints.\\
These studies fall into the following two categories:
\begin{enumerate}
\item The first type of constraints was developed by Gomez-Reino and Scrucca \cite{GomezReino:2007qi,GomezReino:2008px}. Subject to a vanishing cosmological constant, they developed necessary but not sufficient constraints for the existence of vacua in a general supergravity setup consisting of D-term and F-term potential.
\item In more recent papers Covi et al. \cite{Covi:2008ea,Covi:2008cn} studied explicitly the possibilities for various string compactifications to obtain metastable dS minima and inflation. However, the constraints used here do not assume a vanishing cosmological constant but do not include a D-term potential.
\end{enumerate}
In order to avoid a long calculation and introduction of terminology to the reader we simply would like to argue why our approach falls into the category of string models discussed in \cite{Covi:2008ea,Covi:2008cn} and why this also allows us to satisfy the modified constraints after the inclusion of D-term potentials:

First of all, our model heavily relies on both components of the scalar potential, D-term and F-term. Since we are in principle able to tune the minimum of our potential to zero cosmological constant, we assume that we can use the constraints developed in the first series of papers by Gomez-Reino and Scrucca. The constraints we have to satisfy are:
\begin{eqnarray}
 f^i f_i+d^a d_a &=& 1\, ,\\
\nonumber R_{i\bar{j}p \bar{q}}f^i f^{\bar{j}}f^p f^{\bar{q}}&\leq& \frac{2}{3}+ \frac{2}{3} (M_{ab}^2/m_{3/2}^2-2h_{ab})d^a d^b+2h^{cd}h_{aci}h_{bd \bar{j}}f^{i}f^{\bar{j}}d^a d^b\\
&& -(2 h_{ab}h_{cd}-h^i_{ab}h_{cdi})d^a d^b d^c d^d+\sqrt{\frac{3}{2}}\frac{Q_{abc}}{m_{3/2}}d^ad^b d^c\, ,
\end{eqnarray}
where $f_i$ and $d_a$ are the F-term or respectively D-term rescaled by $1/m_{3/2}$ and $R_{i\bar{j}p \bar{q}}$ denotes the Riemann tensor with respect to the K\"ahler metric. $h_{ab}$ is the gauge kinetic function and $Q_{abc}$ is the variation of the gauge kinetic function with respect to the generators of the gauge symmetries.\footnote{A more detailed explanation of the quantities in those constraints and a derivation of those can be found in the original papers.}

Neglecting the D-term for the moment, the second constraint simplifies to
\begin{equation}
 R_{i\bar{j}p \bar{q}}f^i f^{\bar{j}}f^p f^{\bar{q}}\leq \frac{2}{3}\, ,
\end{equation}
 which was renamed in the second series of papers to $\sigma= R_{i\bar{j}p \bar{q}}f^i f^{\bar{j}}f^p f^{\bar{q}}-\frac{2}{3}> 0.$ For various types of string compactifications, the value of $\sigma$ was determined in the work by Covi et al. In particular it was shown that for no-scale models with included $\alpha'-$corrections, it is always possible to satisfy this constraint for a vanishing cosmological constant (cf. equation (4.33) in \cite{Covi:2008ea}).

Including D-terms can even alleviate this problem, as shown in \cite{GomezReino:2007qi,GomezReino:2008px} since one can rescale the F-terms and the curvature in such a way that the dependence on the D-terms seems to disappear:
\begin{eqnarray}
 \delta_{I\bar{J}}z^I z^{\bar{J}}&=&1\, ,\\
\tilde{R}_{I\bar{J}P \bar{Q}}z^{I}z^{\bar{J}}z^{P}z^{\bar{Q}}&\leq& \frac{2}{3}\, ,
\end{eqnarray}
where $z^I=f^I/\sqrt{1-\sum_A d_A^2}.$ The change in the curvature can be evaluated in particular limits of the relation between gaugino and gravitino mass. In the terminology of those papers, we are working in the regime of the ``light vector limit'' since $g_{U(1)} M_{U(1)}/2m_{3/2}< g_{U(1)}\ll 1.$ In this case it was shown in \cite{GomezReino:2007qi} that the curvature is lowered due to the D-terms.\\
Hence we can conclude that in our class of models we are easily able to satisfy the constraints for general supergravity setups and the possibility of generating stable dS minima.

\subsection{Constraints on Anomalous $U(1)$ Gauge Symmetries as Uplifting Potential}
Our explicit construction for dS moduli stabilisation with the help of an anomalous $U(1)$ gauge symmetries 
contrasts with the findings  of Choi and Jeong in \cite{Choi:2006bh}.
In this article they consider the possibility of using the D-term potential of an anomalous $U(1)$ symmetry as the uplifting mechanism and find that no uplifting is possible if the gravitino mass is smaller than the Planck scale by many orders of magnitude, assuming the K\"ahler moduli of order unity. This assumption is clearly violated in the context of the LARGE volume scenario discussed here which then allows to have D-term uplifting. Recall 
also that, unlike the KKLT scenario for which the original F-term vanishes, the LARGE volume scenario has non-vanishing F-terms and therefore non-vanishing D-terms are also possible. 
\section{Conclusions}
In this article we have found a probably unexpected application of the ISS
scenario. Embedding it within low-energy effective actions from type IIB
string compactifications, the original metastable minimum of the ISS
scenario tends to be destabilised towards runaway in the direction of the
K\"ahler modulus determining the gauge coupling constants. This happens
even after introducing a constant flux induced superpotential $W_0$ in the
simplest one-modulus case.

Things change dramatically in Swiss-cheese compactifications with several
K\"ahler moduli.
We have found that in these cases, the effective potential for the
K\"ahler moduli, obtained after matter field stabilisation has a similar
form to the original LARGE volume scenario. Contrary to that case, in
which the non-perturbative superpotential is assumed not to depend on
chiral matter fields, the
scalar potential cannot be minimised analytically. But the similarity
allows for a numerical
treatment that detects an exponentially large volume, with several
differences. The main difference is that the necessary presence of
D-terms, induced by anomalous $U(1)$'s gives a positive definite
contribution to the scalar potential in such a way that well inside the
K\"ahler cone ($\tau_1\gg \tau_2$) the potential goes to zero from above.
Therefore this allows naturally to de Sitter compactifications with
exponentially  large
volume. Using the fluxes, the minimum of the potential can be tuned to
essentially zero value.

This allows for interesting applications: First, for the computation
of soft supersymmetry breaking terms in a realistic context in which
the ISS scenario plays the role of hidden sector and the Standard
Model brane wraps a cycle without non-perturbatively induced
superpotential. The soft terms are universal as in the standard
LARGE volume scenario since the fields that break supersymmetry are
the K\"ahler moduli which unlike the complex structure moduli, are
insensitive to flavour. Generically the soft terms will be all of
order the gravitino mass, but cancelations are possible depending on
the modular weights of the matter fields.

Secondly, our scenario allows also the possibility of realising K\"ahler
moduli inflation and Fibre Inflation in a fully supersymmetric set-up, the
previous realisations were obtained using an ad-hoc uplifting term.
Essentially  we used the fact that we can obtain de Sitter space to
provide the positive contribution to the scalar potential and then use a
different K\"ahler modulus as the inflaton, just as in the original
formulations of both scenarios. A full analysis of the multi-field
system for inflation is out of the scope of this article.

This scenario appears to be quite general. The main fact we used is
the existence of an anomalous $U(1)$ with its D-term potential  and a
nonperurbative contribution to the superpotential of the form
$e^{-aT}\Phi\rho$. Therefore we expect the lifted large volume minimum to
appear in a large class of chiral models for which the non-perturbatively
induced superpotential includes matter fields. An interesting open
question is to embed this scenario within a realistic
compact Calabi-Yau construction.
\section*{Acknowledgements}
We would like to thank Cliff Burgess, Michele Cicoli, Joe Conlon,
Matt Dolan, Marta Gomez-Reino, Anshuman Maharana and Angel Uranga
for useful discussions. SLK is funded by SDW, ETC and EPSRC. FQ is
funded by STFC.
\appendix
\section{On stabilising the F-term potential with a constraint}
In the process of stabilising the D-term and F-term potential separately, we claimed that it is not possible to stabilise all fields in a suitable regime (e.g. at large volume). We now would like to prove this result. The starting point is the schematic potential
\begin{equation}
V=A |\rho \Phi|^2+2 B \text{ Re }(\rho\Phi)+C(\xi+|\rho|^2+|\Phi|^2)\, ,
\end{equation}
where the coefficients depend on all other variables and constants. In addition we have the constraint from the minimisation of the D-term potential.
\begin{equation}
|\rho|^2=x |\Phi|^2+b\, ,
\end{equation}
where $b$ is determined by the FI-contribution to the D-term potential and $x$ denotes the arbitrariness in the charge assignment. Differentiating with respect to $\rho$ and $\Phi$ leads to the following two equations:
\begin{eqnarray}
0&=&A \rho^* |\Phi|^2+B \Phi+C \rho^*\, , \\
0&=&A \Phi^* |\rho|^2+B \rho+C \Phi^*\, .
\end{eqnarray}
Multiplying the first equation by $\rho$ and the second by $\Phi$ and subtracting, this gives $|\Phi|=|\rho|$, which cannot be satisfied since the D-term implies explicitly that they cannot be equal in order to cancel the FI term.

\section{Next to leading order corrections to $\Phi$}
So far we stabilised $\rho$ and $\Phi$ to leading order. Since the stabilisation with respect to $\rho$ depends crucially on the cancellation of the FI term, one can ask what happens if we want to stabilise $\Phi$ up to next to leading order. Does this destabilise the whole stabilisation procedure? In order to answer this question, let us repeat the ansatz for $\Phi:$
\begin{equation}
\Phi=\frac{1}{{\cal V}^{1/6}}\left(\tilde{\Phi}+\frac{\varphi}{{\cal V}^\beta}\right),
\end{equation}
and the next to leading order contribution to $\partial_{\tilde{\Phi}} V$ at the leading order value:
\begin{eqnarray}
\partial_{\tilde{\Phi}}V|_{\tilde{\Phi}=\tilde{\Phi}+\varphi/{\cal V}^\beta}&=&\frac{2}{2\tau_2 {\cal V}^2}\left(\frac{3 \sqrt{\tau_2}\xi}{2a {\cal V}}+\frac{2\tilde{\Phi}\frac{\varphi}{{\cal V}^\beta}(18a\tau_2-n)}{18a \tau_2^{1-n}}\right)\tilde{\Phi}\\
&-&\frac{2}{2\tau_2{\cal V}^{5/2}}\left(\frac{2\tilde{\Phi}(18a\tau_2-n)}{18a \tau_2^{1-n}}\right) \left(\frac{|\tilde{\rho}|^2 \left(36a \tau_2+n\right)}{18a \tau_2^{1-n}}\right)\\
&+&\frac{2\tilde{\Phi}\alpha^2 e^{-2a\tau_2}}{\tau_2^n{\cal V}^{5/3}}\, .
\end{eqnarray}
In our scenario $\varphi$ is taking negative or small values which can be seen as follows: At the LARGE volume minimum, we expect
\begin{equation}
e^{a\tau_2}\sim {\cal V}\, .
\end{equation}
This implies that the contribution proportional to $\tilde{\rho}$ is suppressed with an additional factor $1/{\cal V}^{2/3}$ which makes it subleading compared to the next to leading order corrections coming from the FI bit. Hence, $\beta=1$ and the next to leading order corrections to $\Phi$ at our minimum will be negative. Suppose $e^{a\tau_2}$ is not suppressed with respect to the volume, the contribution from the leading order F-term will be dominating and drives $\varphi$ to negative values again. In an intermediate range the contribution from the cross-term between $\rho$ and $\Phi$ might be leading. However, we were never able to assign $\varphi$ a value to cancel our D-term contribution completely and hence we always keep the uplifting contribution.
\section{Minimizing quark masses}
\subsection*{1-modulus}
In principle there is a flat direction in the leading order D-term potential which allows non-vanishing $q$ and $p.$ So we can minimize the D-term potential for non-vanishing $p,q$ and then we have to try to minimize the F-term potential. Looking at the only non-trivial case of no implicit suppression with respect to the large K\"ahler modulus, the leading order F-term contribution is given by
\begin{equation}
 A |\frac{p\Phi q}{\mu}+\rho \Phi|^2\, ,
\end{equation}
where A contains constants and the dependence on the K\"ahler modulus. The first derivative with respect to $p$ gives
\begin{equation}
 A \Phi q (\frac{p\Phi q}{\mu}+\rho \Phi)^*\, ,
\end{equation}
which is only zero for non-vanishing $\Phi$ and $\rho$ if $q=0$, since $pq/\mu+\rho\neq 0$ due to the rank condition.
The case of an implicit suppression with respect to the K\"ahler moduli is discussed below.
\subsection*{2-moduli}
A priori it is clear that the potential is extremized at $p=q=0$ since the dominating D-term potential is extremized. Suppose this extremum corresponds to a maximum and we could find a minimum closeby. At that minimum $p$ has an implicit dependence on the large K\"ahler modulus. This dependence has to be at least of the suppression of $\Phi$ (i.e. $1/\tau_1^{1/4}$). Schematically the D-term potential looks like
\begin{equation}
 V_D\sim(|\Phi|^2-|\rho|^2-|p|^2-{\rm FI})\, .
\end{equation}
Neglecting $\rho$ for the moment, we can see that there is a flat direction in the D-term potential corresponding to $|\Phi|^2-{\rm FI}=|p|^2.$ Looking at next to leading order effects, i.e. the F-term potential in this case we can ``lift'' this flat direction. Since we are perturbing around our solution presented in the main part $|\Phi|^2-{\rm FI}\approx 0$ and in particular it should be suppressed with respect to the volume at least as much as $|\Phi|.$ Then the leading order contribution from the F-term potential still is given by
\begin{equation}
 \frac{|\Phi|^2 e^{-2a \tau_2}}{\tau_1^{5/2}}\, .
\end{equation}
Hence we do not want $|\Phi|$ to become larger at next to leading order, which implies that $|p|=0$ is a minimum.
\section{Estimating scales}
The masses associated with our fields are given as the square root of the eigenvalues of the matrix
\begin{equation}
K^{-1}_{i\bar{j}}\frac{\partial^2 V}{\partial x^i\partial x^{\bar{j}}}\, .
\end{equation}
To understand the results it is very helpful to estimate the leading order contributions to the K\"ahler metric which we find to be
\begin{equation}
K_{i\bar{j}}(\tau_1,\tau_2,\rho,\Phi)=\left(\begin{matrix}
\frac{3 M_P^2}{4\tau_1^2} & -\frac{27 M_P^2 \tau_2^{7/6}+2 g_s^{-2/3} |\tilde{\Phi}|^2}{24\tau_1^{5/2}\tau_2^{2/3}}&-\frac{g_s^{-2/3}\tilde{\rho}\tau_2^{1/3}}{2\tau_1^{29/12}}&-\frac{g_s^{-2/3}\tau_2^{1/3}\tilde{\Phi}}{2\tau_1^{9/4}}\cr
 -\frac{27 M_P^2 \tau_2^{7/6}+2 g_s^{-2/3}|\tilde{\Phi}|^2}{24\tau_1^{5/2}\tau_2^{2/3}}& \frac{27 M_P^2 \tau_2^{7/6}-4g_s^{-2/3} |\tilde{\Phi}|^2}{72\tau_1^{3/2}\tau_2^{5/3}}&\frac{g_s^{-2/3}\tilde{\rho}}{6\tau_1^{17/12}\tau_2^{2/3}}& \frac{g_s^{-2/3}\tilde{\Phi}}{6\tau_1^{5/4}\tau_2^{2/3}}\cr
-\frac{g_s^{-2/3}\tilde{\rho}^*\tau_2^{1/3}}{2\tau_1^{29/12}}&\frac{g_s^{-2/3}\tilde{\rho}^*}{6\tau_1^{17/12}\tau_2^{2/3}}&\frac{g_s^{-2/3}\tau_2^{1/3}}{\tau_1}&0\cr
-\frac{\tau_2^{1/3}\tilde{\Phi}^*}{2g_s^{2/3}\tau_1^{9/4}}& \frac{\tilde{\Phi}^*}{6g_s^{2/3}\tau_1^{5/4}\tau_2^{2/3}}&0 &\frac{g_s^{-2/3}\tau_2^{1/3}}{\tau_1}\cr
\end{matrix}
\right)
\end{equation}
This leading order behaviour in the K\"ahler metric is with respect to the the K\"ahler moduli almost the same as in the LARGE volume case. The only change is given by the contribution proportional to $\Phi$ in the relevant components but this change is negligibly small.

We find the leading order volume suppression in the Hessian of the potential evaluated at the minimum for the brane moduli to be:
\begin{equation}
d^2 V=\left(\begin{matrix}
\frac{1}{\tau_1^{5}}& \frac{1}{\tau_1^{3+1/3}} & \frac{1}{\tau_1^{4+5/12}} & \frac{1}{\tau_1^{3+3/4}}\cr
\frac{1}{\tau_1^{3+1/3}}& \frac{1}{\tau_1^{3}} & \frac{1}{\tau_1^{3+5/12}} & \frac{1}{\tau_1^{2+3/4}}\cr
\frac{1}{\tau_1^{4+5/12}}& \frac{1}{\tau_1^{3+5/12}} & \frac{1}{\tau_1^{3+5/6}} & \frac{1}{\tau_1^{2+2/3}}\cr
\frac{1}{\tau_1^{3+3/4}}& \frac{1}{\tau_1^{3+3/4}} & \frac{1}{\tau_1^{2+2/3}} & \frac{1}{\tau_1^{2+1/2}}\cr
\end{matrix}\right)
\end{equation}
where we assumed that at $e^{a\tau_2}\sim \tau_1^{3/2}$ and we neglected the exact coefficients for simplicity. Numerically this leads to the following masses for the dS example from the previous section:
\begin{eqnarray}
m_{\tau_1}^2&=&6.7 \times 10^{-7} M_P^2\\
m_{\tau_2}^2&=& 2.5 \times 10^{-6} M_P^2\\
m_{\rho}^2&=& 2.7 \times 10^{-7} M_P^2\\
m_{\Phi}^2&=& 3.9 \times 10^{-5} M_P^2
\end{eqnarray}
In order to interpret these results we would like to compare the masses with the masses in the LARGE volume scenario, which are given by
\begin{eqnarray}
m_{\tau_1}&\sim& \frac{g_s^2 W_0}{({\cal V}_s^0)^{3/2}}M_P\sim m_{3/2}\left(\frac{m_{3/2}}{M_P}\right)^{1/2}\, ,\\
m_{\tau_2}&\sim& \frac{a g_s W_0}{{\cal V}_s^0}M_P\sim 2m_{3/2}\ln{M_P/m_{3/2}}\, .
\end{eqnarray}
Taking now the same parameters as for the dS example, we obtain
\begin{eqnarray}
m_{\tau_1}^2 &=&4.1 \times 10^{-8} M_P^2\, ,\\
m_{\tau_2}^2 &=&1.2 \times 10^{-5} M_P^2\, .
\end{eqnarray}
Compared with the original LARGE volume scenario, we can conclude that we obtain roughly the same masses for the K\"ahler moduli. One of the brane moduli becomes very massive $m_\Phi$ and the other mass $m_{\rho}$ is relatively the lightest particle but it is roughly of the same order as the large K\"ahler modulus $\tau_1.$ The difference in the K\"ahler moduli masses is due to different numerical prefactors in the volume of the Calabi-Yau. In order to compare these results completely reliably, we would have to go to a realistic value of the volume which goes beyond our simple dS example. Before such an analysis is possible we should estimate further crucial properties of this modified model such as the structure of supersymmetry breaking.

\section{An approximative analytic stabilisation of the large K\"ahler modulus}
After integrating out the matter field degrees of freedom we ended up with the following potential
\begin{equation}
V=g_s M_P^4 \left(\frac{A e^{-2 a\tau_2}}{\tau_1^{5/2} \tau_2^{1/6}}-\frac{B e^{-4/3 a \tau_2}\tau_2^{14/9}}{\tau_1^{11/3}}+\frac{C}{\tau_1^{9/2}}\right).
\end{equation}
Changing the suppression with respect to the large K\"ahler modulus $\tau_1$ in the term proportional to $B$ from $11/3$ by $1/6$ to $7/2$ should not affect the structure qualitatively. The change in the minimal value is negligibly small but allows us to minimise the potential analytically.

The modified potential takes the classical form of the volume suppression looking like:
\begin{equation}
V=\frac{g_s M_P^4}{\tau_1^{9/2}}\left(\frac{A e^{-2 a\tau_2}\tau_1^2}{ \tau_2^{1/6}}-B e^{-4/3 a \tau_2}\tau_2^{14/9}\tau_1+C\right)
\end{equation}
Demanding the vanishing of the first derivative of the potential with respect to $\tau_1$ leads to
\begin{eqnarray}
0=\frac{\partial V}{\partial \tau_1}&=&-\frac{g_s M_P^4}{2\tau_1^{11/2}}\left(\frac{5 A e^{-2 a\tau_2}\tau_1^2}{ \tau_2^{1/6}}-7B e^{-4/3 a \tau_2}\tau_2^{14/9}\tau_1+9C\right)\\
\Rightarrow 0&=&\tau_1^2-\frac{7B e^{2/3 a\tau_2}\tau_2^{31/18}}{5 A} \tau_1+\frac{9C e^{2a\tau_2}\tau_2^{1/6}}{5A}\\
&=& \tau_1^2-B' e^{2/3a\tau_2}\tau_1+C' e^{2/3 a\tau_2}.
\end{eqnarray}
Solutions to this equation are given by
\begin{equation}
\tau_1^{(\text{min, max})}=\frac{B'}{2}e^{2/3 a\tau_2}\pm \sqrt{\frac{B'^2}{4}e^{4/3a\tau_2}-C' e^{2a\tau_2}}\, .
\end{equation}
Looking at the leading order behaviour of the overall potential, it is clear that the negative solution will correspond to the minimum and the positive solution to the maximum.
In order to keep to real solutions we have to satisfy the following condition for the discriminant:
\begin{equation}
\frac{B'^2}{4C'}>e^{2/3 a\tau_2}.
\end{equation}
The value of the discriminant is also crucial for the value we can stabilise the small K\"ahler modulus $\tau_2$ at:

Demanding the vanishing of the first derivative with respect to the small K\"ahler modulus leads to the following condition:
\begin{equation}
0=4B e^{2/3 a\tau_2}\tau_2^{31/18}(-7+6a\tau_2)-3A \tau_1(1+12 a\tau_2)\, ,
\end{equation}
which is equivalent to writing
\begin{equation}
0=\frac{20}{21}B' e^{2/3a \tau_2}(-7+6a\tau_2)-\tau_1(1+12 a\tau_2)\, .
\end{equation}
In the large $a\tau_2$ limit this condition simplifies at the minimum to the following equation
\begin{eqnarray}
0&=&-\frac{1}{21} B' e^{2/3 a\tau_2}+2  \sqrt{\frac{B'^2}{4}e^{4/3a\tau_2}-C' e^{2a\tau_2}}\\
\Rightarrow \frac{440}{441}\frac{B'^2}{4C'}&=& e^{2/3 a\tau_2}\, .
\end{eqnarray}
This is consistent with the constraint from above as expected and we obtain the same behaviour in an exact calculation.

Unfortunately, we cannot solve the remaining equations for $\tau_2$ analytically. However, we can be sure that a solution exists if the constraint is satisfied. Furthermore these approximations should help in constructing solutions numerically.

\section{Calculating F-terms}
Using $D_S W=0,$ the F-term associated with the dilaton field $S$ is found to be given by at leading order:
\begin{eqnarray}
\nonumber F^{\bar{s}}&=&e^{\frac{K}{2 M_P^2}}\left( K^{\bar{s}\tau_i} D_{\tau_i} W+K^{\bar{s}\phi_i} D_{\phi_i} W\right)\\
\nonumber &=&\frac{1}{g_s{\cal V}}\left(K^{\bar{s}\tau_i} \partial_{\tau_i} W+\frac{W}{M_P^2}K^{\bar{s}\tau_i} \partial_{\tau_i} K+K^{\bar{s}\phi_i} \partial_{\phi_i} W+ \frac{W}{M_P^2}K^{\bar{s}\phi_i} \partial_{\phi_i} K\right)\\
\nonumber &=&\frac{1}{g_s {\cal V}}\left(K^{\bar{s}\tau_2} \partial_{\tau_2} W+M_P g_s^{3/2} W_0 K^{\bar{s}\tau_1} \partial_{\tau_1} K+K^{\bar{s}\Phi} \partial_{\Phi} W+K^{\bar{s}\rho} \partial_{\rho} W\right)\\
\nonumber &=&\frac{1}{g_s {\cal V}}\left(\frac{18\sqrt{2}M_P g_s^{3/2}W_0 s^{5/2}\xi}{\tau_1^{3/2}}+\frac{36 \sqrt{2} e^{-a\tau_2}g^{1/2+m}s^{5/2}\alpha a\xi \tilde{\rho}\tilde{\Phi}}{\tau_1^{13/6}}\right)\\
&=& \frac{18\sqrt{2} M_P g_s^{1/2}W_0 s^{5/2}\xi}{{\cal V}^2}+O(\frac{1}{{\cal V}^{22/9}})\, ,
\end{eqnarray}
which is the same as in the LVS. The F-terms associated with the K\"ahler moduli are at leading order given by:
\begin{eqnarray}
\nonumber F^{\bar{\tau_1}}&=&e^{\frac{K}{2 M_P^2}}\left( K^{\bar{\tau}_1\tau_i} D_{\tau_i} W+K^{\bar{\tau}_1\phi_i} D_{\phi_i} W\right)\\
\nonumber &=&\frac{1}{g_s\left({\cal V}+\frac{\xi}{g_s^{3/2}}\right)}\left(K^{\bar{\tau}_1\tau_i} \partial_{\tau_i} W+\frac{W}{M_P^2}K^{\bar{\tau}_1\tau_i} \partial_{\tau_i} K+K^{\bar{\tau}_1\phi_i} \partial_{\phi_i} W+ \frac{W}{M_P^2}K^{\bar{\tau}_1\phi_i} \partial_{\phi_i} K\right)\\
\nonumber &=&\frac{1}{g_s\left({\cal V}+\frac{\xi}{g_s^{3/2}}\right)}\left(-2 W_0 M_P g_s^{3/2} \tau_1+K^{\bar{\tau}_1\tau_i} \partial_{\tau_i} W+K^{\bar{\tau}_1\phi_i} \partial_{\phi_i} W\right)\\
 &=&\frac{1}{g_s\left({\cal V}+\frac{\xi}{g_s^{3/2}}\right)}\left(-2 W_0 M_P g_s^{3/2} \tau_1-\frac{6\sqrt{2} W_0 M_P \xi}{\sqrt{\tau_1}}+\frac{4a \alpha g_s^{1/2+m}}{M_P}\tilde{\rho}\tilde{\Phi}\tau_1^{1/3}\tau_2 e^{-a \tau_2}\right),
\end{eqnarray}
where unlike in the LVS the third term is dominating over the second one at the minimum. In the LVS as discussed in \cite{Blumenhagen:2009gk} there is a sub-leading cancellation, which is not present due to a different leading order structure of the potential and the inclusion of the D-term uplifting contribution: Here the second term plus the D-term uplifting contribution (with opposite sign) are of the order of the second one, cf. equation (\ref{dspot2}):
\begin{eqnarray}
\nonumber V&\supset& \frac{2\tau_2^{2n-1} |\tilde{\rho}|^4}{ g_s^{1/3}{\cal V}^{22/9}}+\frac{2\sqrt{6}M_P^3 g_s^{2/3}W_0\alpha\sqrt{6}\tau_2^{5/4-n/2}e^{-a\tau_2}\text{ Re }\tilde{\rho}}{{\cal V}^{22/9}}+\frac{3 M_P^4 W_0^2}{2 {\cal V}^{27/9}}
\left(\frac{\xi}{g_s^{3/2}}+\frac{n\sqrt{\tau_2}}{a}\right).
\end{eqnarray}
With an absent D-term uplifting contribution the third term becomes dominant. However, it is out of the scope of this article to determine the precise volume suppression of the F-term when the second and third term cancel themselves partially.
\begin{eqnarray}
\nonumber F^{\bar{\tau_2}}&=&e^{\frac{K}{2 M_P^2}}\left( K^{\bar{\tau}_2\tau_i} D_{\tau_i} W+K^{\bar{\tau}_2\phi_i} D_{\phi_i} W\right)\\
\nonumber &=&\frac{1}{g_s\left({\cal V}+\frac{\xi}{g_s^{3/2}}\right)}\left(K^{\bar{\tau}_2\tau_i} \partial_{\tau_i} W+\frac{W}{M_P^2}K^{\bar{\tau}_2\tau_i} \partial_{\tau_i} K+K^{\bar{\tau}_2\phi_i} \partial_{\phi_i} W+ \frac{W}{M_P^2}K^{\bar{\tau}_2\phi_i} \partial_{\phi_i} K\right)\\
 &=&\frac{1}{g_s\left({\cal V}+\frac{\xi}{g_s^{3/2}}\right)}\left(-2M_P W_0 g_s^{3/2} \tau_2-\frac{8 a \alpha g_s^{1/2+n}\tilde{\rho}\tilde{\Phi}\tau_1^{5/6}\sqrt{\tau_2}e^{-a\tau_2}}{3 M_P}\right),
\end{eqnarray}
\begin{eqnarray}
\nonumber F^{\bar{\tau_3}}&=&e^{\frac{K}{2 M_P^2}}\left( K^{\bar{\tau}_3\tau_i} D_{\tau_i} W+K^{\bar{\tau}_3\phi_i} D_{\phi_i} W\right)\\
\nonumber &=&\frac{1}{g_s\left({\cal V}+\frac{\xi}{g_s^{3/2}}\right)}\left(K^{\bar{\tau}_3\tau_i} \partial_{\tau_i} W+\frac{W}{M_P^2}K^{\bar{\tau}_3\tau_i} \partial_{\tau_i} K+K^{\bar{\tau}_3\phi_i} \partial_{\phi_i} W+ \frac{W}{M_P^2}K^{\bar{\tau}_3\phi_i} \partial_{\phi_i} K\right)\\
&=&\frac{1}{g_s\left({\cal V}+\frac{\xi}{g_s^{3/2}}\right)}\left(-2 M_P W_0 g_s^{3/2} \tau_3- \frac{4a \alpha g_s^{1/2+m}\tilde{\rho}\tilde{\Phi}\tau_2\tau_3 e^{-a\tau_2}}{M_P \tau_1^{2/3}}\right).
\end{eqnarray}
Although it is not of importance in the analysis of the soft-term structure, we would like to mention the fact that there is no cancellation in the F-term associated with $\tau_2$ as in the LVS. This is due to the fact that the leading order contributions in the F-term potential come from various F-terms and not only from the small modulus $\tau_2.$
\providecommand{\href}[2]{#2}\begingroup\raggedright\endgroup

\end{document}